\documentclass{elsart}

\usepackage{graphicx}
\usepackage{amssymb}

\begin{document}

\begin{frontmatter}

\title{Spherical Colloids: Effect of Discrete Macroion Charge Distribution and Counterion
Valence}
\author{Ren\'{e} Messina\corauthref{rene}}
\ead{messina@mpip-mainz.mpg.de}
\corauth[rene]{Tel: +49 6131 379 148  Fax: +49 6131 379 100}
\address{Max-Planck-Institut f\"{u}r Polymerforschung, Ackermannweg 10, 55128 Mainz,
Germany}
\date{\today{}}
\maketitle
\begin{abstract}
We report the coupled effects of macroion charge discretization and counterion
valence in the primitive model for spherical colloids. Instead of considering
a uniformly charged surface, as it is traditionally done, we consider a more
realistic situation where \textit{discrete monovalent microscopic charges} are
randomly distributed over the sphere. Monovalent or multivalent
counterions ensure global electroneutrality. We use molecular dynamics simulations
to study these effects at the ground state and for finite temperature.
The ground state analysis concerns the counterion structure and \textit{charge
inversion}. Results are discussed in terms of simple analytical models. For
finite temperature, strong and weak Coulomb couplings are treated. In this
situation of finite temperature, we considered and discussed the phenomena
of ionic pairing (pinning) and
unpairing (unpinning).\\
\end{abstract}
\end{frontmatter}
%


\section{Introduction}

Charged colloidal suspensions are a subject of intense experimental and theoretical
work not only because of their direct application in industrial or
biological processes, but also because they represent model
systems for atomistic systems. 
The electrostatic interactions involved in such systems have a fundamental
role in determining their physico-chemical properties 
\cite{Isralachvili_1992,Evans_book_1999}.
Theoretical description of highly charged colloidal solutions  faces two challenges:
(i) different typical length scales  due to the presence of macroions
(i. e. charged colloids of the size 10-1000 \AA)\ and microscopic counterions
and (ii) their \textit{long-range} Coulomb interaction.
A first simplifying assumption is to treat the solvent as a dielectric
medium solely characterized by its relative permittivity $ \epsilon _{r} $.
A second widely used approximation consists in modeling the \textit{short range}
ion-ion excluded volume interaction by hard spheres. These two approximations
are the basis of the so-called primitive model of electrolyte solutions. The
system under consideration is an asymmetrical electrolyte solution made up of highly
charged macroions and small counterions. A further simplification motivated
by this asymmetry
can be made by partitioning the system into subvolumes (cells), each containing
one macroion together with its neutralizing counterions plus (if present) additional
salt. This approximation has been called accordingly the cell model
\cite{Hill_book_1960,Wennerstroem_JCP_1982}.
The cells adopt the symmetry of the macroion, here spherical, and are electrostatically
decoupled. It is within the cell model that we present our simulation results.

For spherical macroions the structural charge is usually modeled by a \textit{central}
\textit{charge}, which by Gauss theorem is equivalent to  a \textit{uniform}
surface charge density as far as the electric field (or potential) \textit{outside}
the spherical colloid is concerned.

Most analytical concepts as well as simulations rely on the above assumptions
and especially on the central charge assumption. It is well known that in the
strong Coulomb coupling regime ion-ion correlations become very important, and
significant deviations from mean-field approaches are expected. A counter-intuitive
effect which classical mean-field theories (like Poisson-Boltzmann model) cannot
explain is the phenomenon of \textit{overcharge}, also called charge inversion.
That is, there are counterions in \textit{excess} in the vicinity of the macroion
surface so that its net charge changes sign. This has recently attracted significant
attention \cite{Perel_Physica_1999,Shklowskii_PRE_1999b,Mateescu_EPL_1999,Joanny_EPJB_1999,Sens_PRL_1999,Marcelo_PRE_RapCom1999,Deserno_Macromol_2000,Messina_PRL_2000,Messina_EPL_2000,Nguyen_PRL_2000,Nguyen_JCP_2000,Messina_EPJE_2001,Messina_PRE_2001}.
In particular, we showed recently that this phenomenon may give rise to a strong
long range attraction between like-sign charged colloids
\cite{Messina_PRL_2000,Messina_EPL_2000,Messina_PRE_2001}.
A natural question which comes up is: does overcharge and more generally ion-ion
correlations strongly depend on the way the macroion structural charge is represented
(i. e. uniformly charged or discrete charges on its surface)? In a recent paper
\cite{Messina_EPJE_2001}, we studied such a situation in the strong Coulomb
coupling regime where the macroion charge was carried by divalent microions in the
presence of divalent counterions (\textit{same} ionic valence). In
Ref. \cite{Messina_EPJE_2001} we reported  the
important result showing that overcharge is still possible under those
conditions. Moreover we showed that the intrinsic electric field solely
due to the macroion surface microions (without counterions) varies strongly
from point to point on the colloidal sphere \cite{Messina_EPJE_2001}.

The goal of this paper is to study by means of molecular dynamics
(MD) simulations the \textit{coupled} effects of macroion charge
discretization and counterion valence in the primitive model for
spherical colloids. A systematic comparison with the uniform
macroion charge distribution (i. e. central charge) is undertaken. The
paper is organized as follows. In Sec. \ref{sec.Simu-model} we
give some details on the macroion charge discretization as well as
on the MD simulation model. Section \ref{sec.GS} is devoted to the
ground state analysis where surface counterion structure and
overcharge are addressed. In Sec. \ref{sec.Temp} we investigate
the finite temperature situation, where counterion structure is
studied for strong and weak Coulomb couplings. Finally, in
Sec. \ref{sec.Conclusion} we provide a summary of the results.

\section{Simulation model\label{sec.Simu-model}}

\subsection{Macroion charge discretization\label{sec.simu-DCC}}

\begin{figure}
\begin{center}
\includegraphics[width = 6 cm]{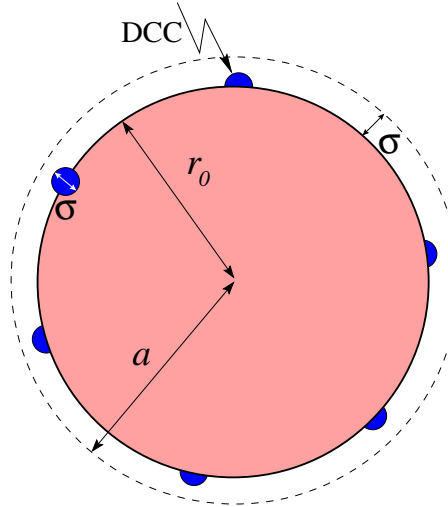}
\end{center}
\caption{
Schematic view of the setup: the discrete colloidal charges (DCC) of diameter
$ \sigma  $ are in dark grey. For a detailed meaning of the
other symbols see text. Note that this a a two-dimensional representation of
the three-dimensional system.
}
\label{fig.setup}
\end{figure}

The procedure is similar to the one used in a previous study \cite{Messina_EPJE_2001}.
The $discrete$ macroion charge distribution is produced by using $Z_{m}$ $monovalent$
microions of diameter $\sigma$ (same diameter as  the counterions) distributed
\textit{randomly} on the surface of the macroion. Then the structural charge
is $Q=-Z_{m}e=-Z_{m}Z_{d}e$, where $ Z_{m}>0 $, $ Z_{d}=1 $ is the
valence of these discrete microions and $e$ is the positive elementary
charge.
These discrete colloidal charges (DCC) are
\textit{fixed} on the surface of the spherical macroion.
Figure \ref{fig.setup} shows a schematic view of the setup.
The counterions (not shown in Fig. \ref{fig.setup}) have a charge $q=+Z_{c}e$, where $Z_{c}>0$
stands for the counterion valence.
In spherical coordinates
the elementary surface is given by:
\begin{equation}
\label{eq.dA}
dA=r_{0}^{2}\sin\theta d\theta d\varphi =-r_{0}^{2}d(\cos\theta )d\varphi \: ,
\end{equation}
where $r_0$ is the distance between the macroion center and the DCC
center. Thus to produce a random discrete charge distribution on the surface
we generated (uniformly) randomly the variables $\cos\theta$ and $\varphi$.
Excluded volume is taken into account by rejecting configurations leading to
an overlap of microions.  Phenomena such as surface chemical reactions \cite{Spalla_JCP_1991},
hydration, roughness \cite{Bhattacharjee_Lang_1998} are not considered. For
commodity we introduce the notation $ (-Z_{d}:+Z_{c}) $ to characterize the
valence symmetry (asymmetry) for $Z_{c}=1$ ($Z_{c}>1$) of the
ions (DCC and counterions) involved in  discrete systems.

\subsection{Molecular dynamics procedure\label{sec.simu-details}}

A MD simulation technique was used to compute the motion
of the counterions coupled to a heat bath acting through a weak stochastic force
\textbf{W}(t). The procedure is very similar to the one used in previous studies
\cite{Messina_PRL_2000,Messina_EPJE_2001}.

The motion of counterion \textit{i} (DCC ions being $fixed$) 
obeys the Langevin equation 
%
\begin{equation}
\label{eq.Langevin}
m\frac{d^{2}{\mathbf{r}}_{i}}{dt^{2}}=
-\nabla _{i}U({\mathbf{r}}_{i})-m\gamma \frac{d{\mathbf{r}}_{i}}{dt}+{\mathbf{W}}_{i}(t)\: ,
\end{equation}
%
where \textit{m} is the counterion mass, \textit{U} is the potential force
having two contributions: (i) the Coulomb interaction and (ii) the excluded volume interaction,
and $ \gamma  $ is the friction coefficient. Friction and stochastic force
are linked by the dissipation-fluctuation theorem
$ <{{\mathbf{W}}_{i}}(t)\cdot {{\mathbf{W}}_{j}}(t')>=6m\gamma k_{B}T\delta _{ij}\delta (t-t^{'}) $.
For the ground state simulations the stochastic force is set to zero.

Excluded volume interactions are taken into account with a pure repulsive Lennard-Jones
potential given by

\begin{equation}
\label{eq. LJ}
U_{LJ}(r)=\left\{ \begin{array}{l}
4\epsilon_{LJ}
\left[ \left( \frac{\sigma }{r-r_{0}}\right) ^{12}-
\left( \frac{\sigma }{r-r_{0}}\right) ^{6}\right] +\epsilon_{LJ} ,\\
0,
\end{array}\qquad \right. \begin{array}{l}
\textrm{for}\, \, r-r_{0}<r_{cut},\\
\textrm{for}\, \, r-r_{0}\geq r_{cut},
\end{array}
\end{equation}
%
where $ r_{0}=0 $ for the microion-microion interaction (the microion being
either a counterion or a DCC), $ r_{0}=7\sigma  $ for the macroion-counterion
interaction and $r_{cut}=2^{1/6}\sigma$ is the cutoff radius.
This leads to a (center-center) macroion-counterion distance of closest approach $ a=8\sigma  $
(see also Fig. \ref{fig.setup}). The macroion surface charge density $ \sigma _{m} $
is defined as

\begin{equation}
\label{eq.sigma-M}
\sigma _{m}=\frac{Z_{m}}{4\pi a^{2}}.
\end{equation}

Energy and length units in our simulations are related to experimental units
by taking $ \epsilon_{LJ} = $$ k_{B}T_{0} $ (with $ T_{0}=298 $ K) and
$\sigma = 3.57 \mathrm{\AA}$  respectively.

The pair electrostatic interaction of any pair \textit{ij}, where \textit{i}
and \textit{j} denote either a DCC a counterion or the central charge (for the
non-discrete case), reads

\begin{equation}
\label{eq.coulomb}
U_{coul}(r)=k_{B}T_{0}l_{B}\frac{Z_{i}Z_{j}}{r}\: ,
\end{equation}
where $ l_{B}=e^{2}/4\pi \epsilon _{0}\epsilon _{r}k_{B}T_{0} $ is the Bjerrum
length describing the electrostatic strength.
For the rest of this paper, electrostatic energy
will always be expressed in units of $k_BT_0$. This also holds for the ground
state analysis where the temperature is $T=0$ K but $T_0=298$ K. From now on
the pair electrostatic interaction will be written in reduced units so that 
Eq. (\ref{eq.coulomb}) reads $U_{coul}=Z_iZ_j/r$. 

The macroion and the counterions are confined in a spherical impenetrable cell
of radius \textit{R.} The macroion is held fixed and is located at the center
of the cell. The colloid volume fraction \textit{$ f_{m} $} is defined as
$a^{3}/R^{3}$. To avoid image charge complications, the permittivity $\epsilon _{r}$
is supposed to be identical within the whole cell (including the macroion) as
well as outside the cell. Typical simulation parameters are gathered in Table
\ref{tab.simu-param}.

\begin{table}[t]
\caption{
Simulation parameters with some fixed values.
}
\label{tab.simu-param}
\begin{tabular}{ll}
\hline
 parameters&
\\
\hline
$ \sigma =3.57 $ \AA\ &
 Lennard Jones length units\\
 $ T_{0}=298K $&
 room temperature\\
 $ \epsilon _{LJ}=k_{B}T_{0} $&
 Lennard Jones energy units\\
 $ Z_{m} $&
 macroion valence\\
$ Z_{d}=1 $&
discrete colloidal charge valence\\
 $ Z_{c} $&
 counterion valence\\
 $ l_{B} $&
 Bjerrum length\\
 $ a=8\sigma  $&
 macroion-counterion distance of closest approach \\
$ \sigma _{m} $&
macroion surface charge density\\
$ R=40\sigma  $&
simulation cell radius\\
 $ f_{m}=8\times 10^{-3} $&
 macroion volume fraction\\
\hline
\end{tabular}
\end{table}

\section{Ground state analysis\label{sec.GS}}

In this section, we focus on counterion distribution exclusively governed by
\textit{energy minimization}, i. e. \textit{T} = 0K. In such a case correlations
are maximal and all the counterions lie on the macroion surface.
This situation has the advantage to enable accurate computation of energy variations
in processes such as overcharging and also to provide a clear description of effects which are
purely correlational in nature. The method employed here was successfully carried
out in Refs. \cite{Messina_PRL_2000,Messina_EPL_2000,Messina_EPJE_2001,Messina_PRE_2001}
and is explained in details in Ref. \cite{Messina_PRE_2001}. The Bjerrum length
$l_{B}$ is set to $10\sigma$. Note that in the ground state the value
of $l_{B}$, or equivalently the value of the dielectric constant $ \epsilon _{r} $,
does not influence at all the counterion structure. Only the electrostatic energy
is rescaled accordingly.

\subsection{Neutral case\label{sec.GS-neutral}}

First we consider the simple case where the system {[}macroion + counterions{]}
is globally neutral. In order to characterize the two-dimensional counterion
structure we compute the counterion correlation function (CCF) $g_c(r)$
on the surface of the sphere defined as

\begin{equation}
\label{eq.CCF-gr}
c^{2}g_c(r=|r'-r''|)=\sum _{i\neq j}\delta (r'-r_{i})\delta (r''-r_{j}),
\end{equation}
%
where $ c=N_{c}/4\pi a^{2} $ is the surface counterion concentration ($ N_{c}=Z_{m}/Z_{c}$
being the number of counterions) and $ r $ corresponds to the arc length
on the sphere. Note that at zero temperature all equilibrium configurations
are identical (except for degenerate ground state), thus
only one is required to obtain $g_c(r)$. The
counterion correlation function $g_c(r)$ is normalized as follows

\begin{equation}
\label{eq.gr-normalization}
c\int _{0}^{\pi a}2\pi rg_c(r)dr=(N_c-1) .
\end{equation}
Because of the \textit{finite} size and the topology of the sphere, $ g(r) $
has a cut-off at $ r_{gc}=\pi a=25.1\sigma  $ and $ g(r_{gc})=0 $. 
Furthermore the
absolute value of $ g(r) $ can not be directly compared to the one obtained
with an infinite plane. 

Similarly, one can also define a surface
macroion correlation function (MCF) $g_m(r)$ for the microions (representing
the colloidal structural charge) on the surface of the macroion.
The normalization of  $g_m(r)$ is very similar to
Eq. (\ref{eq.gr-normalization}) and reads

\begin{equation}
\label{eq.gr_m-normalization}
\sigma_m\int _{0}^{\pi a}2\pi rg_m(r)dr=(Z_m-1) ,
\end{equation}
where  the arc length has been rescaled by a factor $ a/r_{0} $ so that
$g_c(r)$  and $g_m(r)$ are directly comparable (see also the setup
Fig. \ref{fig.setup}) and are defined in the same $r$ range.

\subsubsection{Monovalent counterions\label{sec.GS-neutral-monov}}

\begin{figure}
\begin{center}
\includegraphics[width = 6.8 cm]{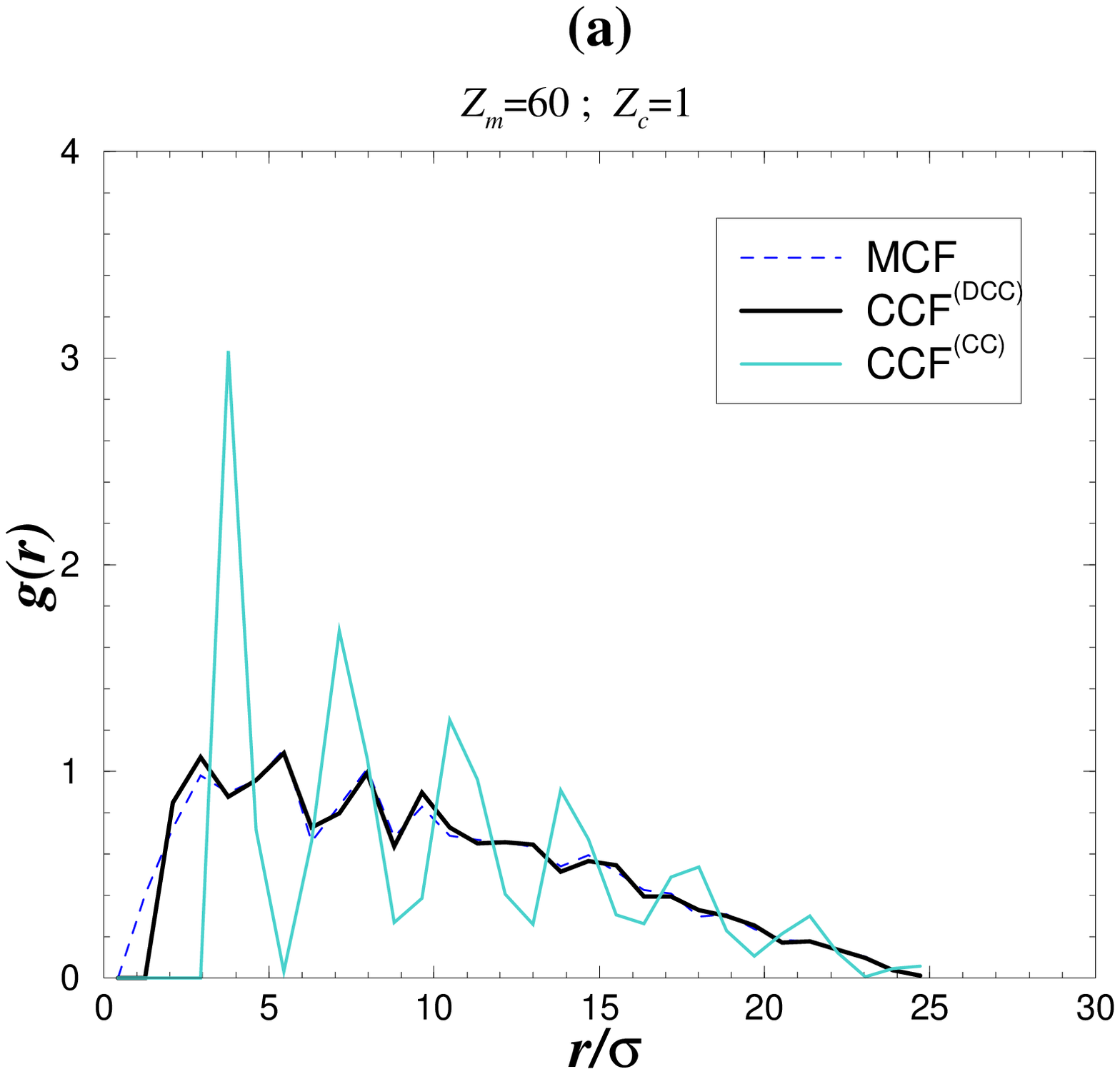}
\includegraphics[width = 6.8 cm]{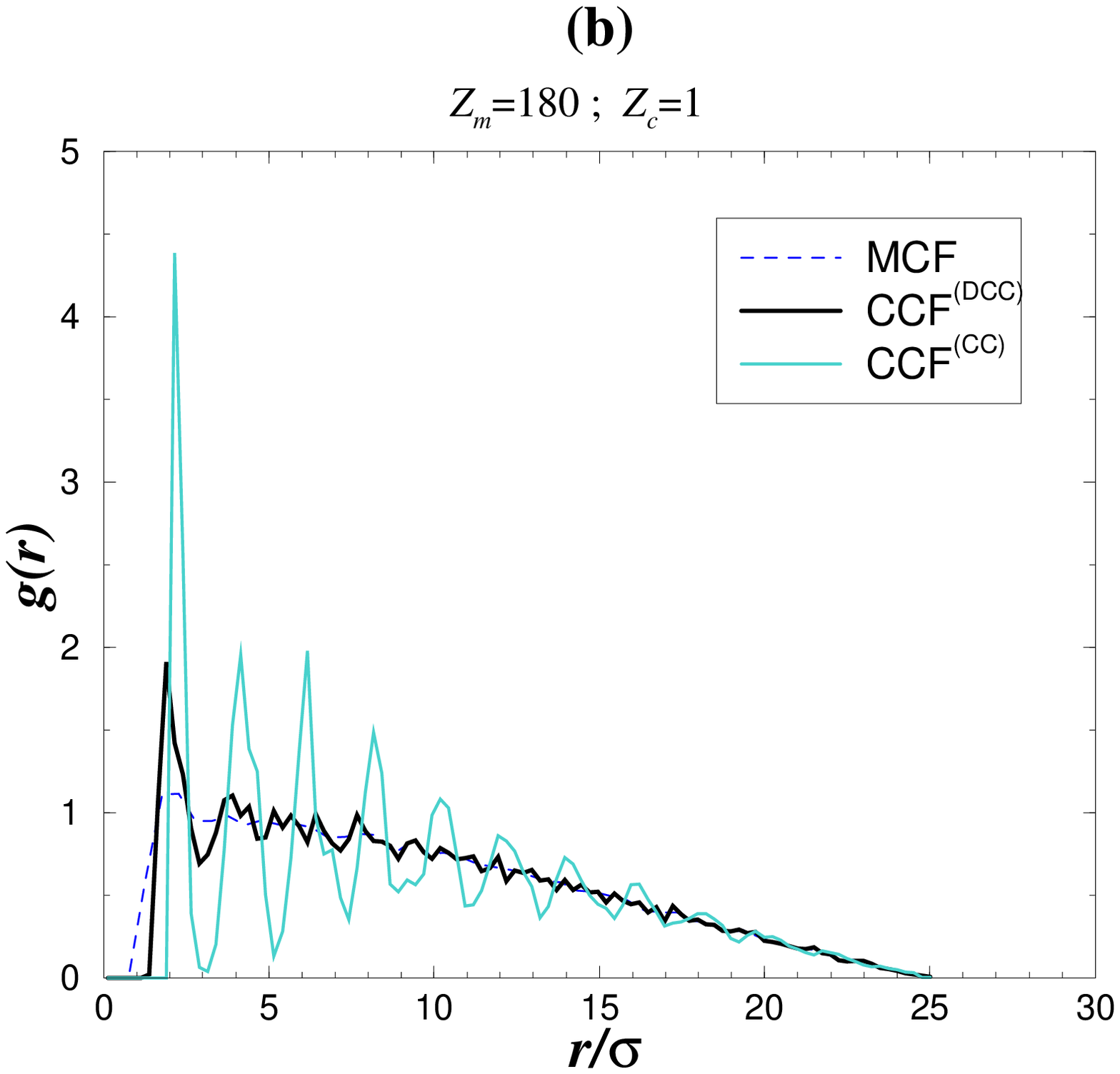}
\includegraphics[width = 6.8 cm]{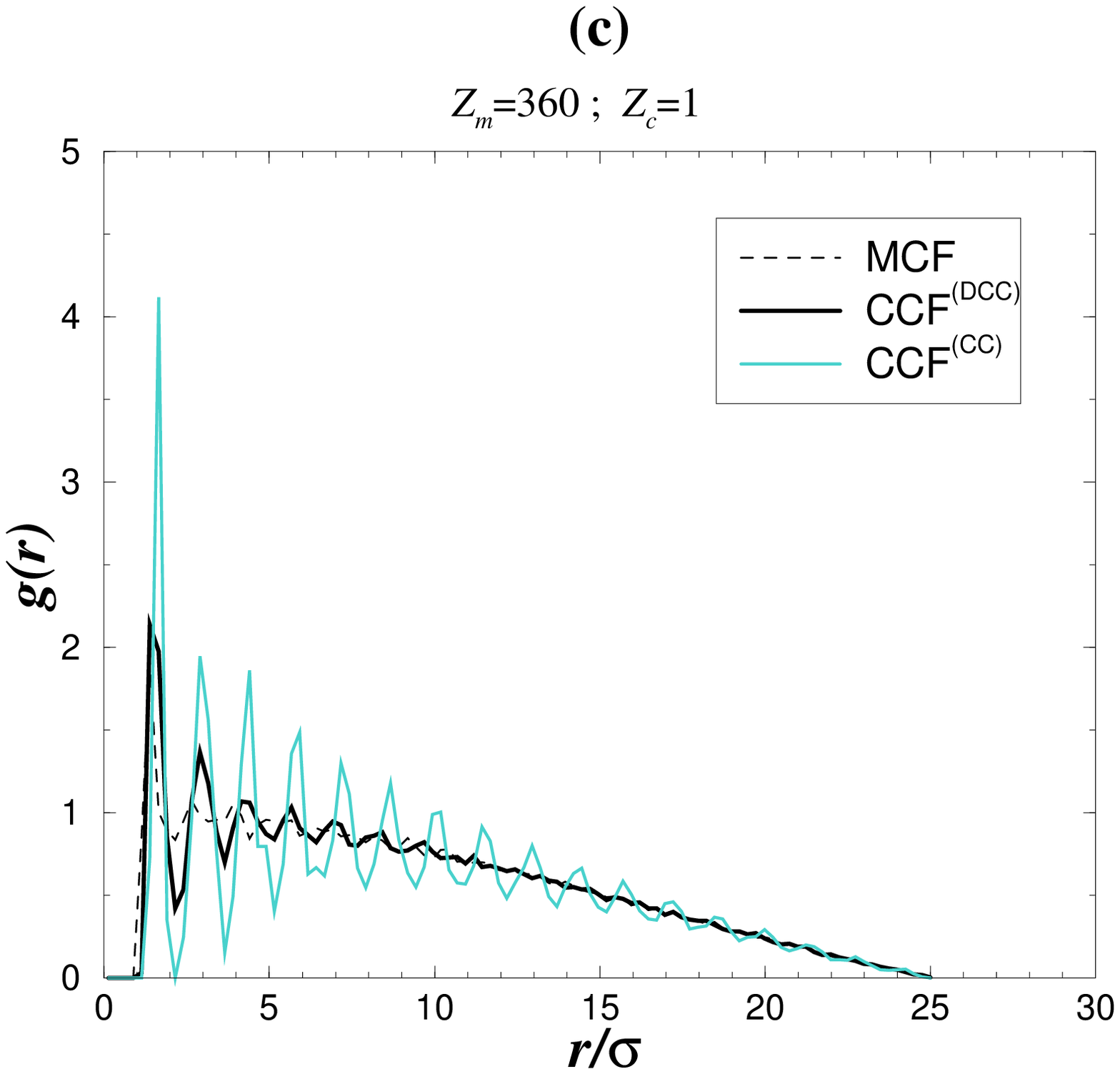}
\hfill
\end{center}
\caption
{
Ground state surface correlation functions $ g(r) $ for
\textit{monovalent} counterions ($ Z_{c}=1 $) and for three
macroion bare charges: (a) \textit{$ Z_{m}=60 $} (b) \textit{$ Z_{m}=180 $}
and (c) \textit{$ Z_{m}=360 $}. The two counterion correlation
functions (CCF) $g_c(r)$ are obtained for discrete colloidal charges
[$g_c^{ \mathrm {(DCC)}}(r)$  denoted by $\mathrm {CCF}^{ \mathrm {(DCC)}}$]
and for the central charge
[$g_c^{ \mathrm {(CC)}}(r)$ denoted by $\mathrm {CCF}^{ \mathrm {(CC)}}$].
MCF stands for the discrete colloidal charges pair distribution $g_m(r)$.
}
\label{fig.gs-CCF-monovalent}
\end{figure}

We first treat the systems where we have monovalent counterions, that is we
have to deal with the symmetric discrete system (-1:+1). The counterion
correlation functions $ g_c(r) $ are computed for a central macroion charge
[denoted by $g_c^{ \mathrm {(CC)}}(r)$] and for discrete
macroion charge distribution [denoted by $g_c^{ \mathrm {(DCC)}}(r)$].
Results for three structural charges $ Z_{m}=60 $,
180 and 360 are given in Figs. \ref{fig.gs-CCF-monovalent}(a), (b) and (c)
respectively. For the continuous case (central charge) the counterion structure 
consists of a pseudo-Wigner crystal
(WC) as was already found in Refs.
\cite{Messina_PRL_2000,Messina_EPL_2000,Messina_EPJE_2001,Messina_PRE_2001}.
Also the higher the absolute number of counterions $ N_{c} $ (i. e. the concentration
$ c $) the higher the order of counterion structure for the continuous case
{[}compare Fig. \ref{fig.gs-CCF-monovalent}(a) with Fig. \ref{fig.gs-CCF-monovalent}(c){]}.

\begin{figure}
\begin{center}
\includegraphics[width = 6.8 cm]{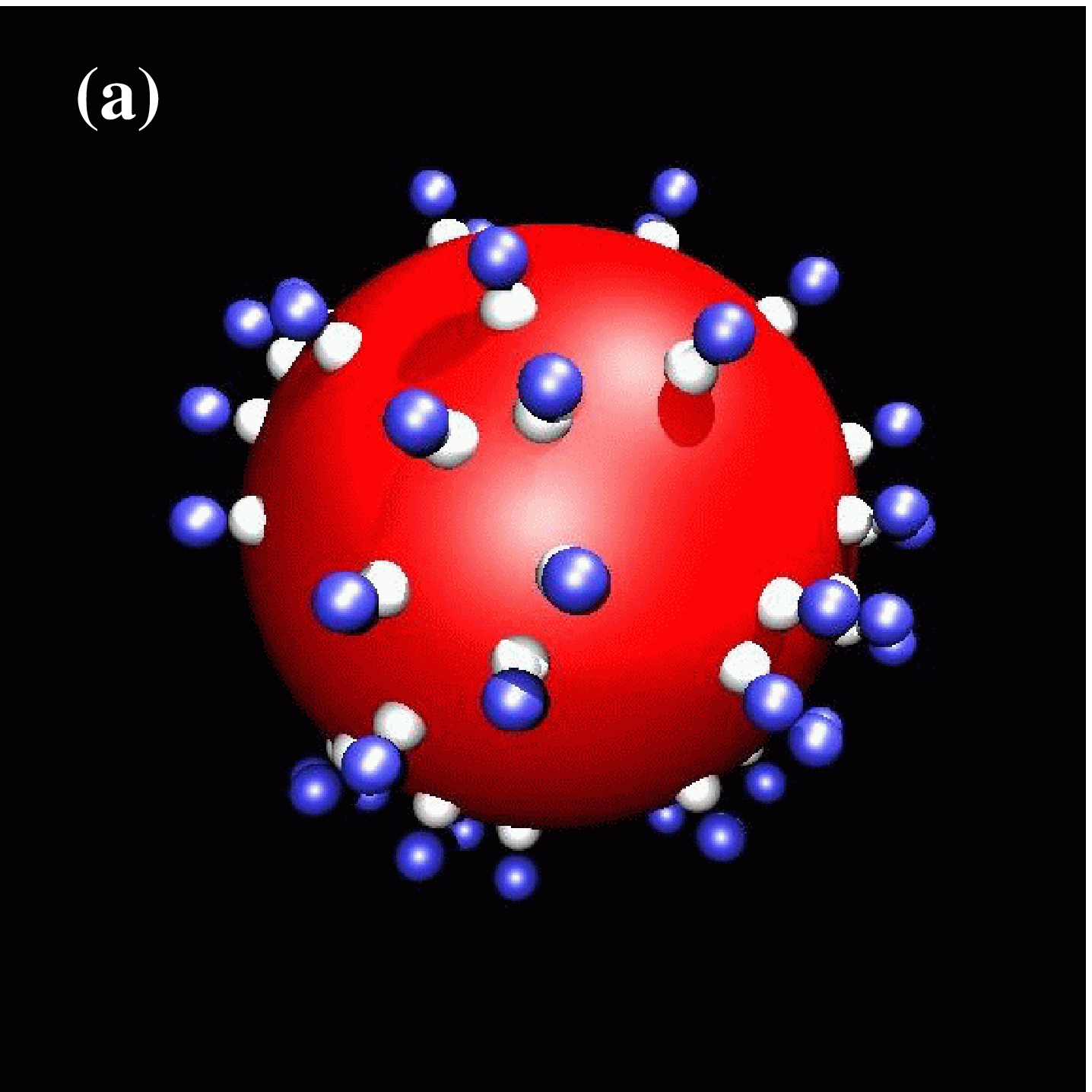}
\includegraphics[width = 6.8 cm]{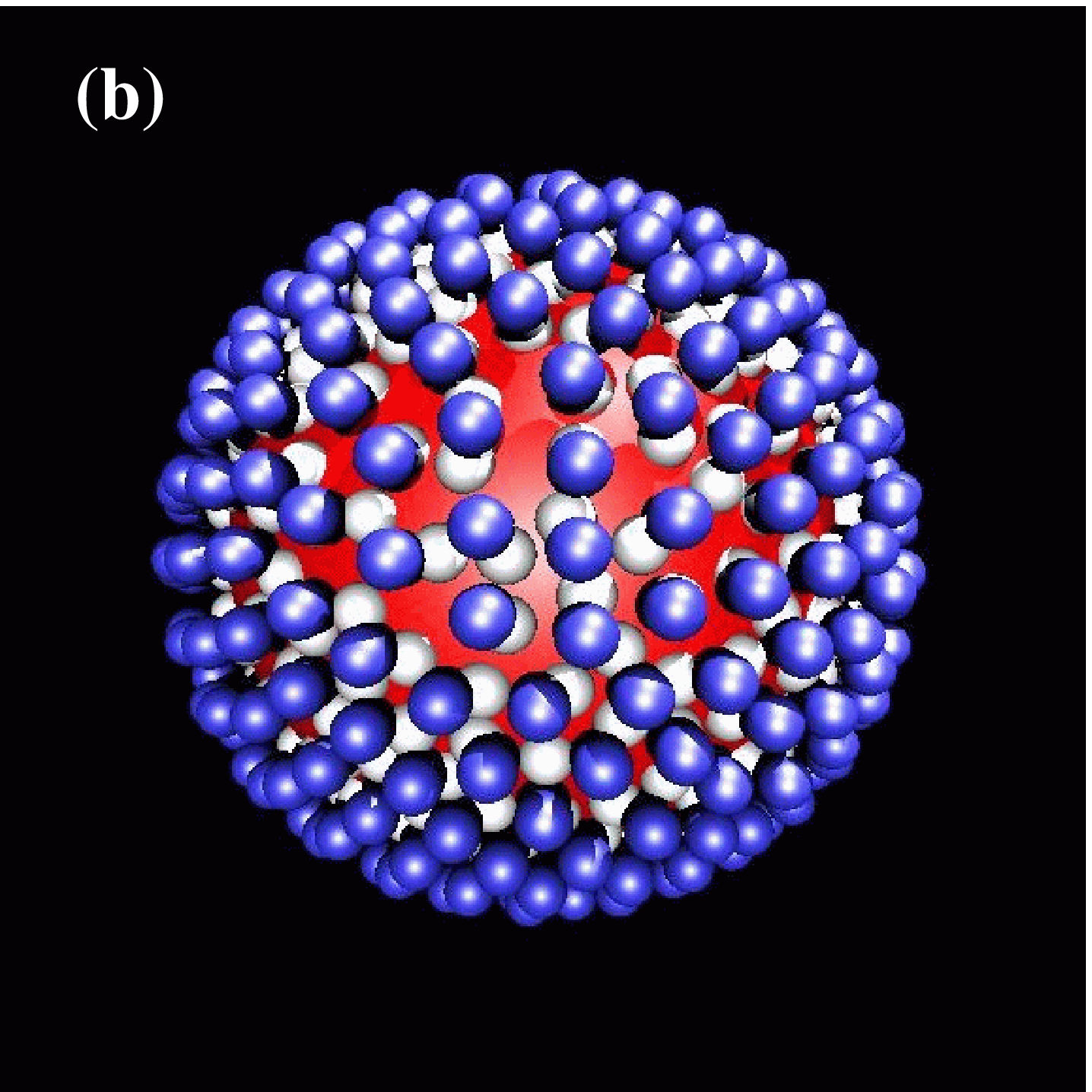}
\hfill
\end{center}
\caption{
Ground state structures for discrete monovalent systems (-1:+1): (a) $ Z_{m}=60 $
and (b) $ Z_{m}=360 $. The discrete colloidal charges
(DCC) are in white, and the counterions in blue. Full ionic pairing association
occurs. The corresponding counterion correlation functions $g_c(r)$
can be found in Figs. \ref{fig.gs-CCF-monovalent}(a) and (c).
}
\label{fig.gs-snapshot-mono-Zm-60-360}
\end{figure}

It turns out that in the case of discrete colloidal charges the counterion
distribution is strongly dictated by the colloidal charge distribution
and especially for low macroion surface charge density $ \sigma _{m} $
($ Z_{m}=60 $) {[}see Fig. \ref{fig.gs-CCF-monovalent}(a){]}. For $ Z_{m}=60 $,
$g_c^{ \mathrm {(DCC)}}(r)$ and $g_m(r)$ are almost identical.
This indicates that each counterion is exactly associated with one
DCC site. The ground state structure for $ Z_{m}=60 $ is depicted in 
Fig. \ref{fig.gs-snapshot-mono-Zm-60-360}(a) where one clearly observes this ionic pairing.

\begin{figure}[b]
\begin{center}
\includegraphics[width = 7 cm]{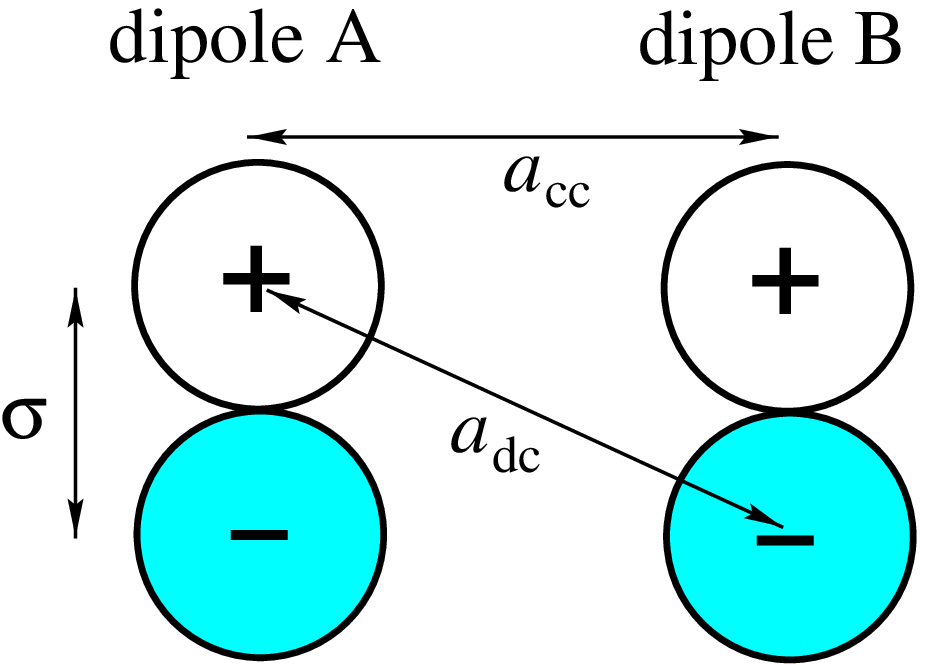}
\hfill
\end{center}
\caption
{
Schematic view of the local electrostatic interactions and typical correlation
lengths involved between nearest dipoles. The negatively charged DCC (-) are
in grey and the positively charged counterions (+) in white.
}
\label{fig.dipoles}
\end{figure}

This strong ionic pair association can be easily explained in terms of \textit{local}
correlations. Let us consider the picture sketched in Fig. \ref{fig.dipoles}
which holds for strong ionic pairing, where a given dipole A (ionic pair made
up of a counterion and a DCC site) on the macroion surface essentially interacts
with its first nearest surrounding dipoles B. Note that very similar lengths
were also considered in a recent theoretical study \cite{Schmitz_1999} in the one-dimensional case
(counterion adsorption on a linear polyelectrolyte).
%
%
It is important to have in mind
that such a local description is physically justified due to the strong screening
generated by ionic pairing. Thereby local correlations are twofold: (i) the
\textit{attractive} interaction between the DCC site of dipole A with its paired
counterion and the counterions of dipoles B, and (ii) the \textit{repulsive}
interaction between the counterion of dipole A and counterions of dipoles B.
The correlations between DCC sites are not relevant since they are fixed. The
intra-dipole attractive interaction $ E_{pin} $ between the DCC site and
its {}``pinned{}'' counterion can be written as

\begin{equation}
\label{eq.Epin}
E_{pin}=-\frac{Z_{d}Z_{c}}{\sigma }.
\end{equation}
For the elementary nearest inter-dipole (or inter-ionic pair) interactions,
one can write for the attractive interaction $ E_{+-} $ between the DCC
site of dipole A and the counterion of dipole B:

\begin{equation}
\label{eq.E+-}
E_{+-}=-\frac{Z_{d}Z_{c}}{a_{dc}}.
\end{equation}
 A similar expression can be written for the repulsive inter-dipole interaction
$ E_{++} $ involving counterions of dipole A and dipole B, which reads

\begin{equation}
\label{eq.Erep}
E_{++}=\frac{Z^{2}_{c}}{a_{cc}}.
\end{equation}
%
Note that the repulsive counterion-counterion term $ E_{++} $ alone, even
if space \textit{truncated}\footnote{%
This statement holds if the cutoff is larger than the lattice constant.
}, drives to the \textit{long-range} ordered triangular WC structure.
However at zero
temperature the DCC sites represent pinning centers for the counterions where
the electrostatic potential is considerably lowered (due to the term $ E_{pin} $)
compared to its direct {}``empty{}'' neighborhood (charge hole), which in
turn prevents the counterions from adopting the ideal WC structure. This latter
aspect was thoroughly discussed in Ref. \cite{Messina_EPJE_2001}. Another important
quantity characterizing discrete systems is the ratio

\begin{equation}
\label{eq.d_pin}
\rho _{pin}=\frac{\widetilde{a}_{cc}}{d_{pin}}
\end{equation}
%
between the mean inter-dipole separation $ \widetilde{a}_{cc} $ (more
exactly the mean counterion-counterion separation) and intra-dipole 
separation $ d_{pin} $ of an ionic pair (in the present study $ d_{pin}=\sigma  $
as depicted in Fig. \ref{fig.dipoles}). The value of $ \widetilde{a}_{cc} $
can be obtained by taking the first peak position of $g_c^{ \mathrm {(CC)}}(r)$.

Obviously, for sufficiently low macroion surface charge density $ \sigma _{m} $
(i. e. large $ \rho _{pin} $) the ionic pairing term $ E_{pin} $ will
be dominant and strong ionic pairing occurs. More specifically, when
the typical inter-dipole distance is large compared to the intra-dipole distance
then dipole-dipole interactions are weak (i. e., $|E_{+-}-E_{++}|\ll |E_{pin}|$)
and the DCC distribution dictates the counterion structure. This is what qualitatively
explains our simulation findings for $ Z_{m}=60 $ {[}see Fig. \ref{fig.gs-CCF-monovalent}(a)
and Fig. \ref{fig.gs-snapshot-mono-Zm-60-360}(a){]}.

When  $\sigma_m$ becomes sufficiently
important the situation may become qualitatively different. In this case
dipoles approach each other and because of excluded volume\footnote{%
Note that in the present model no surface dipole flip is allowed which should
also be the case experimentally.
} $ a_{cc} $ becomes comparable to $ \sigma  $ (see Fig. \ref{fig.dipoles})\footnote{%
The limiting case is where the global structure is compact, i. e. touching spherical
microions.
}. Thereby, the counterion-counterion repulsion term $ E_{++} $ (overcompensating
$ E_{+-} $) induces counterion ordering 
\textit{compatible} with the local attractive pinning potential field generated
by DCC centers. This effect can be inspected in Fig. \ref{fig.gs-CCF-monovalent}(b)
and Fig. \ref{fig.gs-CCF-monovalent}(c) where one sees that upon increasing
$ \sigma _{m} $, $g_c^{ \mathrm {(DCC)}}(r)$ is gradually less correlated
with $g_m(r)$ and more correlated with $g_c^{ \mathrm {(CC)}}(r)$. As a topological
consequence, some counterions will be in contact with several (two or more)
DCC attractors as can be seen in Fig. \ref{fig.gs-snapshot-mono-Zm-60-360}(b).

The quasi-triangular counterion arrangement for high $ \sigma _{m} $ ($ Z_{m}=360 $)
can be inspected in Fig. \ref{fig.gs-snapshot-mono-Zm-60-360}(b). For this symmetric
system in size (same diameter for the counterions and the DCC ions) one expects
that for a compact amorphous DCC layer the counterion structure should become
perfectly ordered. This extreme limit which would correspond to
unreachable experimental charge densities has not been addressed in our simulations.

In parallel, increasing $ \sigma _{m} $ induces by purely excluded volume
effect a stronger local order within the DCC layer as can be checked on the
$g_m(r)$ plots in Figs. \ref{fig.gs-CCF-monovalent}(a)-(c). This is quite similar
to what occurs in a system of hard spheres where the (dense) liquid phase is
locally correlated and the (dilute) gaseous phase is uncorrelated.

In summary, the system depicted above is the siege of an \textit{order-disorder}
phase transition where upon increasing $ \sigma _{m} $ (i. e. decreasing
$ \rho _{pin} $) we pass from a disordered counterion structure (imposed by the DCC
layer) to a \textit{long-range} ordered one which is induced by \textit{local} counterion-counterion
correlations.

Although results presented  above concern one given random distribution
(for each $ Z_{m} $), we carefully checked that similar results and conclusions
could be drawn for different random realizations (systematically five). This
also holds for the following section below where we deal with multivalent counterions.

\subsubsection{Multivalent counterions\label{sec.GS-neutral-multiv}}

\begin{figure}
\begin{center}
\includegraphics[width = 6.5 cm]{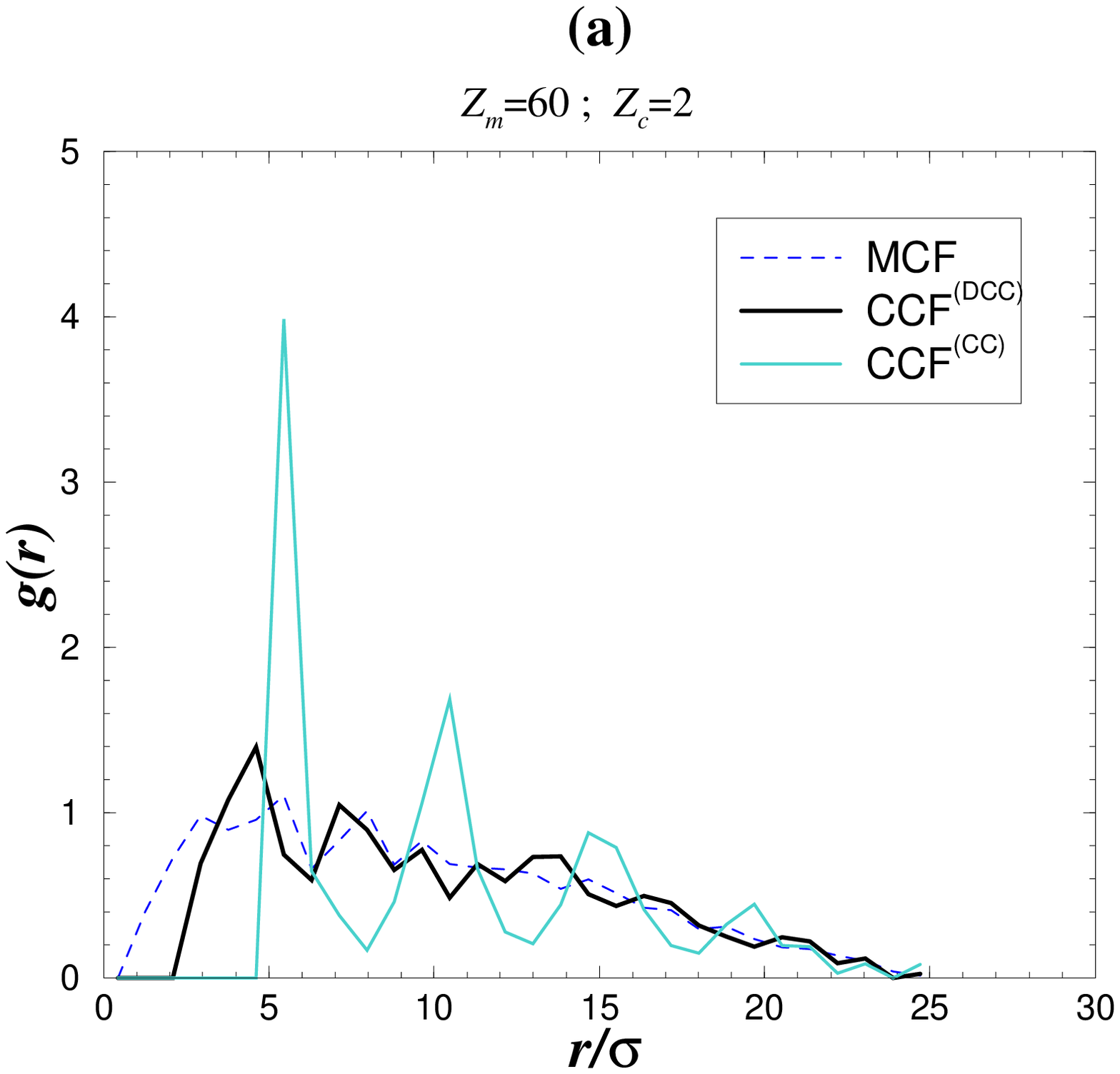}
\includegraphics[width = 6.5 cm]{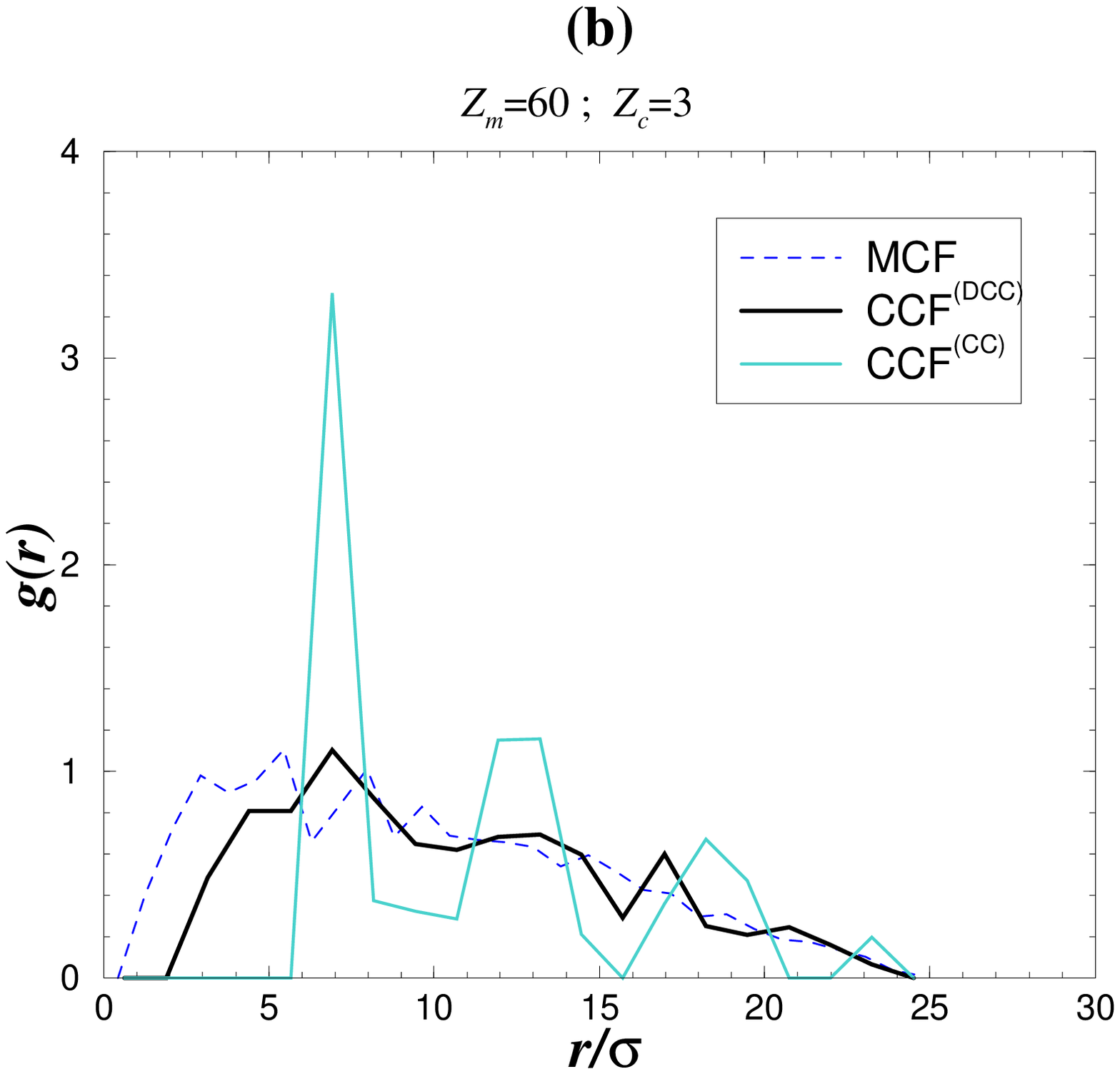}
\includegraphics[width = 6.5 cm]{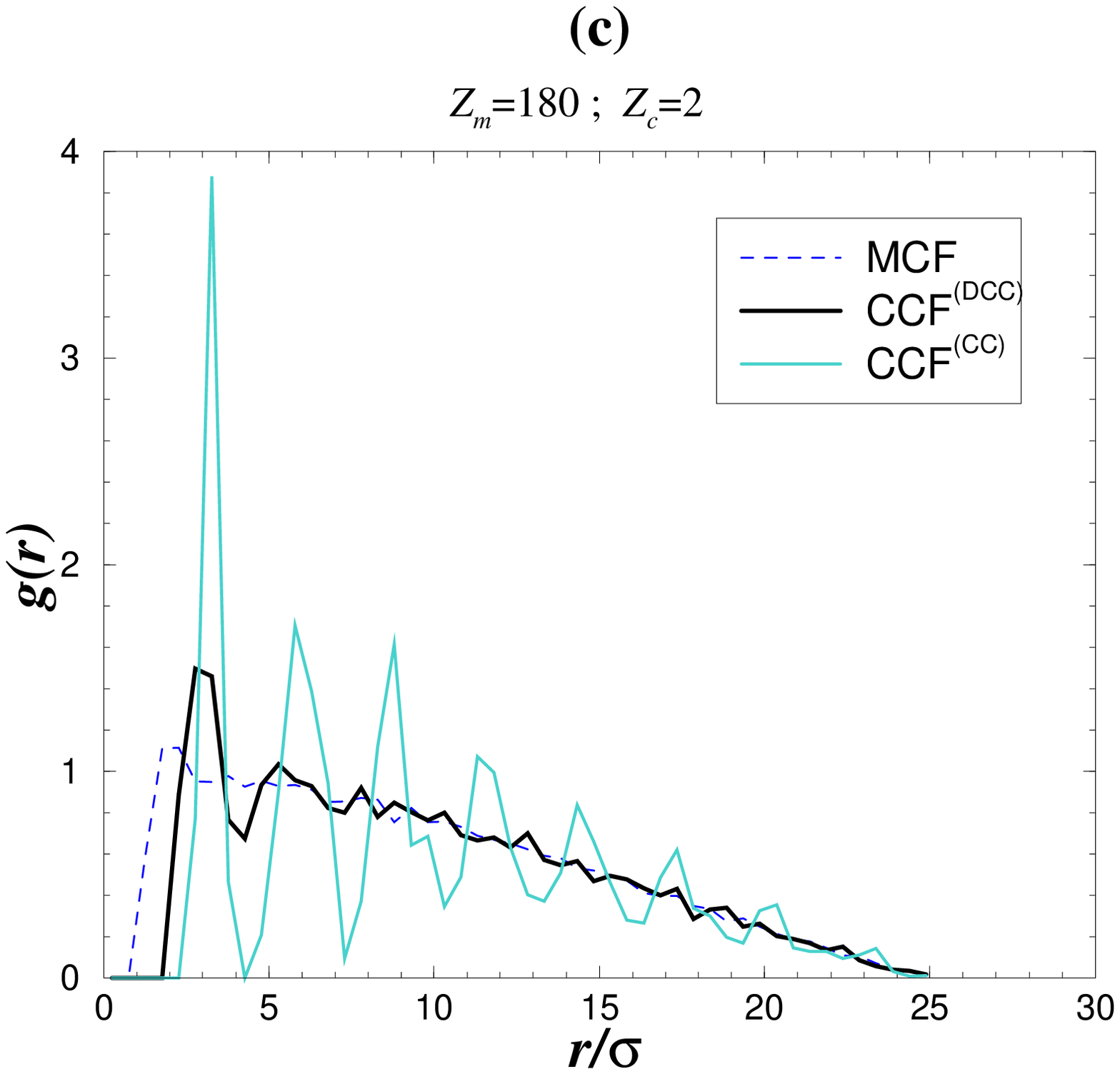}
\includegraphics[width = 6.5 cm]{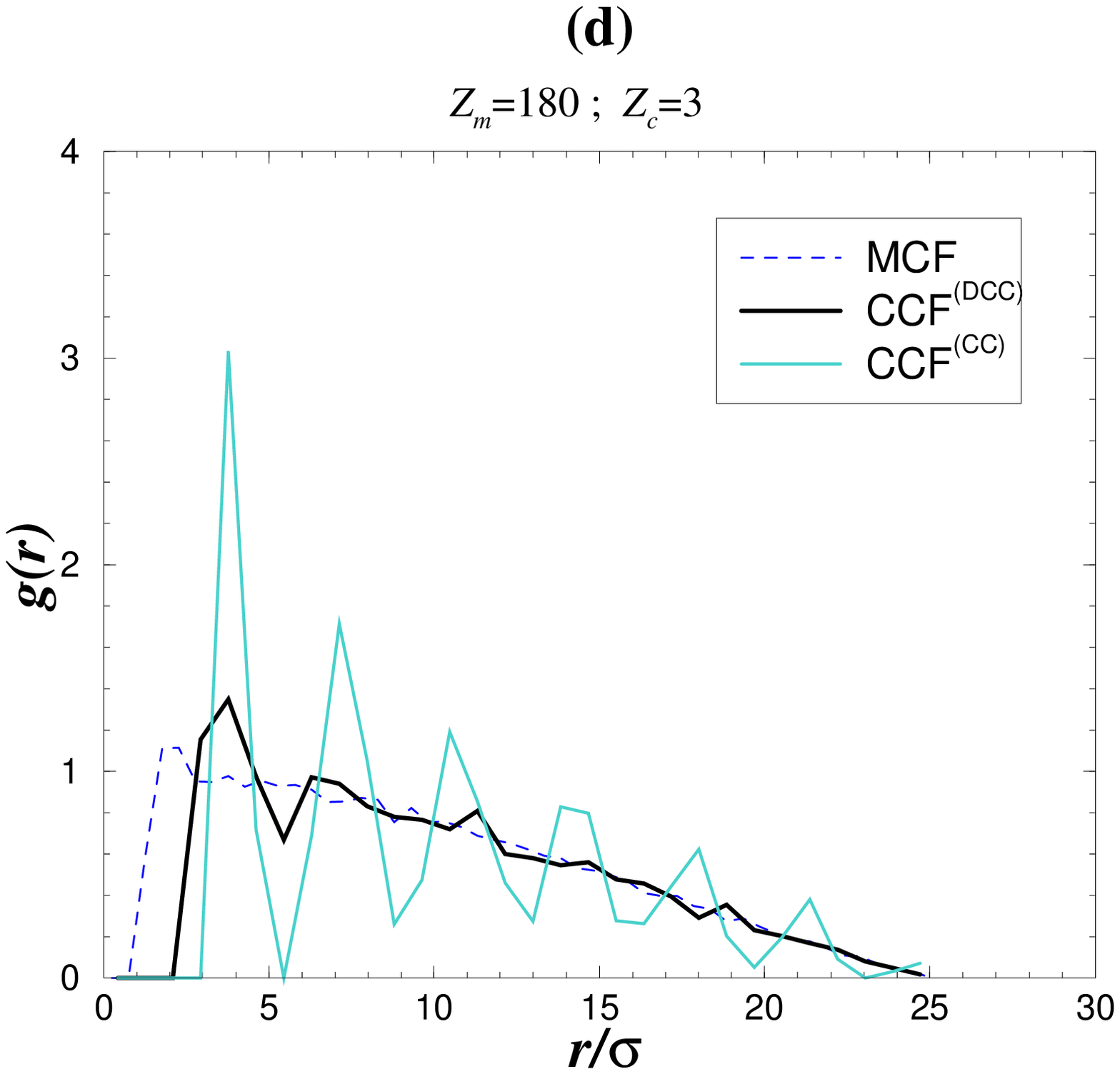}
\includegraphics[width = 6.5 cm]{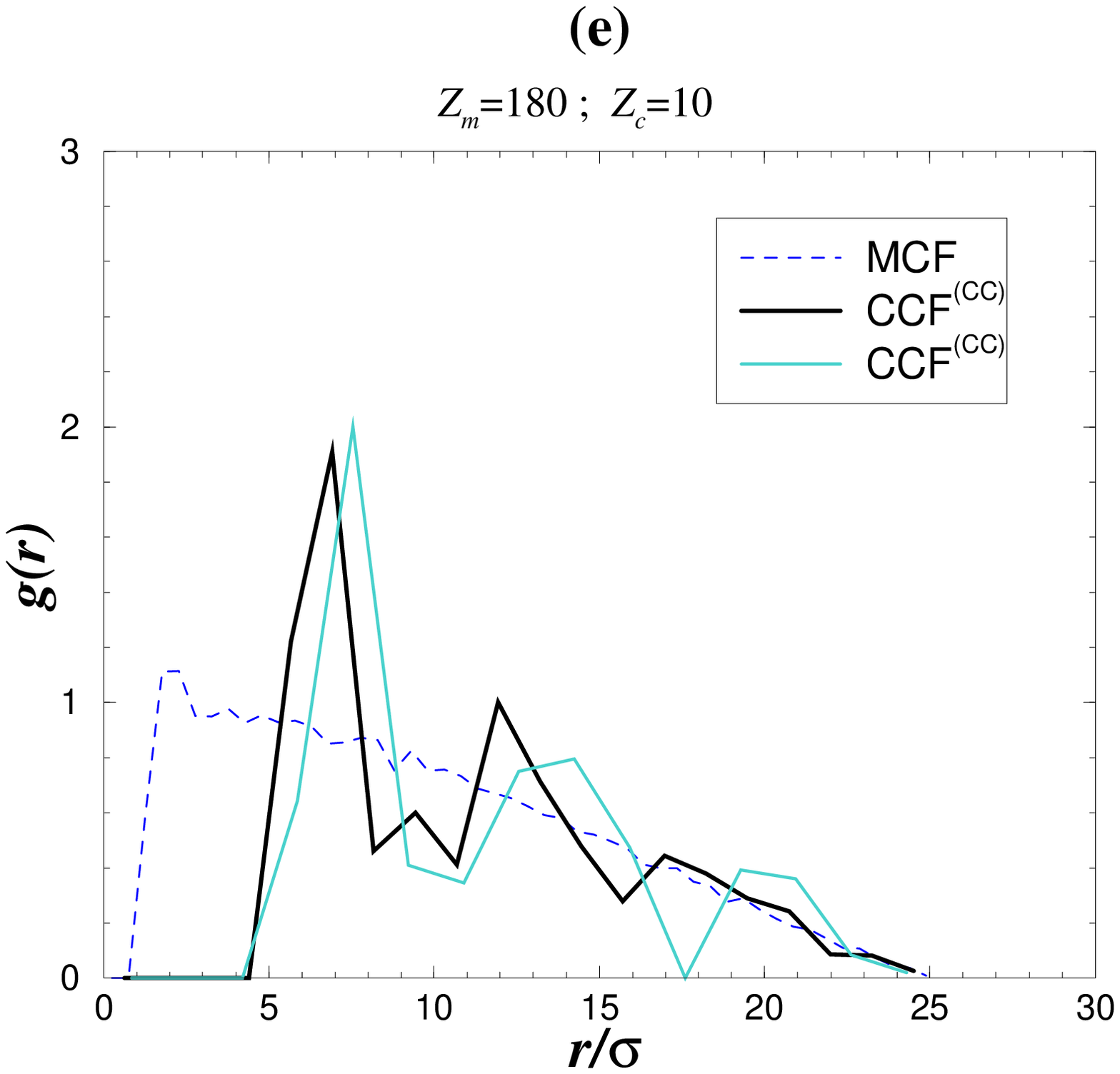}
\hfill
\end{center}
\caption
{
Ground state surface correlation functions for different \textit{multivalent}
systems: (a) \textit{$ Z_{m}=60 $,} $ Z_{c}=2 $;
(b) \textit{$ Z_{m}=60 $,} $ Z_{c}=3 $;
(c) \textit{$ Z_{m}=180 $,} $ Z_{c}=2 $;
(d) \textit{$ Z_{m}=180 $,} $ Z_{c}=3 $;
(e) \textit{$ Z_{m}=180 $}, $ Z_{c}=10 $.
The two counterion correlation functions (CCF) are obtained for discrete colloidal
charges (DCC) and for the central charge (CC). The   $g_m(r)$ curves (denoted by MCF)
are identical from (a) to (b) and from (c) to (e).
}
\label{fig.gs-CCF-multivalent}
\end{figure}

We turn to the asymmetric discrete systems $ (-1:+Z_{c}) $ where multivalent
counterions are present ($ Z_{c}>1 $). The correlation functions $ g(r) $
for two macroion charges $ Z_{m}=60 $ and $ Z_{m}=180 $ and various counterion
valences $ Z_{c} $ can be found in Fig. \ref{fig.gs-CCF-multivalent}. One
remarks that upon decreasing the number of counterions $ N_{c} $ (i.e., increasing
$ Z_{c} $) for fixed $ Z_{m} $, the first peak of $g_c(r)$ is gradually shifted
to the right (compare also the monovalent case given in Fig. \ref{fig.gs-CCF-monovalent})
whatever the nature of the macroion charge is (discrete or continuous). 
Furthermore,
we observe for the discrete systems that upon increasing $ Z_{c} $ (for fixed
$Z_{m}$) the correlation between $g_c^{ \mathrm {(DCC)}}(r)$ and
$g_m(r)$ decreases and increases between $g_c^{ \mathrm {(DCC)}}(r)$ and
$g_c^{ \mathrm {(CC)}}(r)$. This effect is clearly noticeable in Fig.
\ref{fig.gs-CCF-monovalent}(b) and Figs. \ref{fig.gs-CCF-multivalent}(c-e)
corresponding to $ Z_{m}=180 $. The very high counterion valence $ Z_{c}=10 $
reported in Fig. \ref{fig.gs-CCF-multivalent}(e) was undertaken in order to
stress the counterion multivalence effect. These findings lead to the 
conclusion that the counterion valence has the effect of reducing the disorder
in the counterion structure stemming from the randomness of the DCC distribution.

This related phenomenon can be theoretically explained with simple ideas. Basically,
the mechanisms involved in this counterion valence induced ordering stem from
two concomitant sources: (i) \textit{topological} and (ii) correlational.

\begin{figure}
\begin{center}
\includegraphics[width = 7 cm]{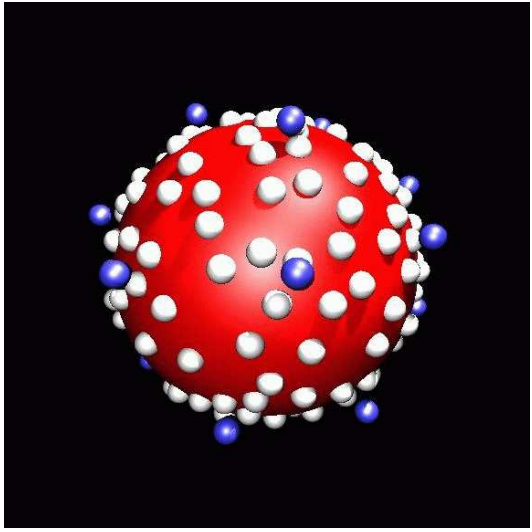}
\end{center}
\caption
{
Ground state structure for (-1:+10) with $Z_{m}=180$.
The corresponding counterion correlation function $g_c(r)$
can be found in Figs. \ref{fig.gs-CCF-multivalent}(e).
}
\label{fig.gs-snapshot-multi-Zm180}
\end{figure}

The topological aspect is due to the presence of $(Z_{m}-Z_{m}/Z_{c})$
unbound DCC sites (free of associated counterion) ensuring global electroneutrality
{[}compare for instance Fig. \ref{fig.gs-snapshot-mono-Zm-60-360}(a) and Fig.
\ref{fig.gs-snapshot-multi-Zm180}{]}. 
It is
to say that here, compared to the monovalent case (-1:+1), the counterions have
all the more {}``freedom{}'' to choose their pinning locations
because $ Z_{c} $ is high. To be more precise, the number of topologically
accessible {}``pinned{}'' configurations is given
\footnote
{
Rigorously, Eq. (\ref{eq.C}) holds when each counterion is associated with
one and only one DCC site (case of low $ \sigma _{m} $). For high $ \sigma _{m} $,
it remains a good approximation to capture the essential physics.
}
by

\begin{equation}
\label{eq.C}
C^{\frac{Z_{m}}{Z_{c}}}_{Z_{m}}=\frac{Z_{m}!}
{\left( Z_{m}-\frac{Z_{m}}{Z_{c}}\right) !\left( \frac{Z_{m}}{Z_{c}}\right) !}
\end{equation}
%
which reduces to 1 for $ Z_{c}=1 $. In the ground state, counterions will
{}``decide{}'' to choose among these various possible arrangements the one
which minimizes the total energy of the system. It is clear that this topological
feature by itself promotes counterion valence induced ordering.

Concomitantly, there is a purely counterion correlation induced ordering which
is $ Z_{c} $ dependent. Indeed, using similar arguments as those previously
employed for monovalent systems (-1:+1) built on Eqs. (\ref{eq.Epin}-\ref{eq.Erep}),
one can infer the role of $ Z_{c} $. More specifically, by assuming an ordered
WC structure\footnote{%
From a topological point of view, it consists in replacing the current (random)
Voronoi structure by the ordered WC structure.
} the term $ E_{++} $ given by Eq. (\ref{eq.Erep}) can be rewritten as

\begin{equation}
\label{eq.E-WC}
E^{WC}_{++}\sim \frac{Z^{3/2}_{c}Z^{1/2}_{m}}{a},
\end{equation}
where $ a_{cc} $ in Eq. (\ref{eq.Erep}) is now given by

\begin{equation}
\label{eq.a_cc}
a_{cc}=\widetilde{a}_{cc}\sim c^{-1/2}\sim \left( \frac{Z_{m}}{Z_{c}a^{2}}\right) ^{-1/2}.
\end{equation}
Equation (\ref{eq.E-WC}) shows that for fixed $ Z_{m} $ and $ a $ (i.
e., fixed macroion charge density) $ E^{WC}_{++} \sim Z^{3/2}_{c} $  whereas
$E_{pin} \sim Z_{c} $ (recalling that $ Z_{d}=1 $)
and therefore for sufficiently high $Z_c$ the term $ E^{WC}_{++}$ will be dominant.
Thereby $ Z_{c} $ induces
counterion ordering so as to minimize  mutual counterion-counterion repulsion merely dictated
by Eq. (\ref{eq.E-WC}). 
As a topological consequence, some
counterions which would be in contact with several DCC sites if they
were monovalent can now be in contact with less DCC sites 
(see Fig. \ref{fig.gs-snapshot-multi-Zm180}).

In summary, these discrete multivalent systems are again the siege of an order-disorder
phase transition which is counterion valence controlled.

\subsection{Overcharge\label{sec.GS-OC}}

We now investigate the charge inversion (overcharge) phenomenon. The starting
equilibrium configurations correspond to neutral ground states as were previously
obtained. The method employed here is very similar to the one used in Refs.
\cite{Messina_PRL_2000,Messina_EPJE_2001}. To produce a controlled overcharge,
one adds successively overcharging counterions (OC) in the vicinity of the macroion
surface. Thereby the resulting system is no longer neutral. Using Wigner crystal
concepts \cite{Shklowskii_PRE_1999b,Bonsall_PRB_1977}, we showed that the gain
in electrostatic energy (compared to the neutral state) by overcharging a single
\textit{uniformly} charged macroion (i.e., central charge) with $n$ overcharging
counterions can be written in the following way
\cite{Messina_PRL_2000,Messina_EPL_2000,Messina_PRE_2001}:

\begin{equation}
\label{Eq.oc}
\Delta E^{OC}_{n}=\Delta E_{n}^{cor}+\Delta E_{n}^{mon} =
-\frac{\alpha Z^{2}_{c}}{\sqrt{A}}\left[ (N_{c}+n)^{3/2}-N^{3/2}_{c}\right] +
Z_{c}^{2}\frac{n^2}{2a}.
\end{equation}
As before
$ N_{c}=Z_{m}/Z_{c} $ is the number of counterions in the neutral state,
$ A $ is the macroion area ($ 4\pi a^{2} $) and $\alpha$ is a positive
constant which was determined by using simulation data for $ \Delta E^{OC}_{1} $.
$ \Delta E_{n}^{cor} $, which is equal to the first term of the right member,
denotes the gain in energy due to ionic correlations. The derivation of this
term can be found in Refs. \cite{Messina_PRL_2000,Messina_EPL_2000,Messina_PRE_2001},
and the basic idea is that each counterion interacts essentially with its neutralizing
uniformly charged Wigner-Seitz cell. The second term on the right hand side,
$ \Delta E_{n}^{mon} $, is the self-energy of the excess of charge.
This repulsive term stops the
overcharging for  sufficiently large $ n $. Note that the WC concept for describing
energy correlations is already excellent for highly short range ordered structures
(strongly correlated liquids, see Ref. \cite{Shklowskii_PRE_1999b} for a detailed
discussion). The total electrostatic energy of the system as a function of $ n $
is displayed Fig. \ref{fig.gs-OC-mono} (for monovalent counterions) and Fig.
\ref{fig.gs-OC-multi} (for multivalent counterions) for two bare charges $ Z_{m}=60 $
and $ Z_{m}=180 $. The energy curves corresponding to discrete systems were
produced by systematically averaging over five random DCC realizations. 

\subsubsection{Monovalent counterions\label{sec.GS-OC-monov} }

\begin{figure}
\begin{center}
\includegraphics[width = 6.8 cm]{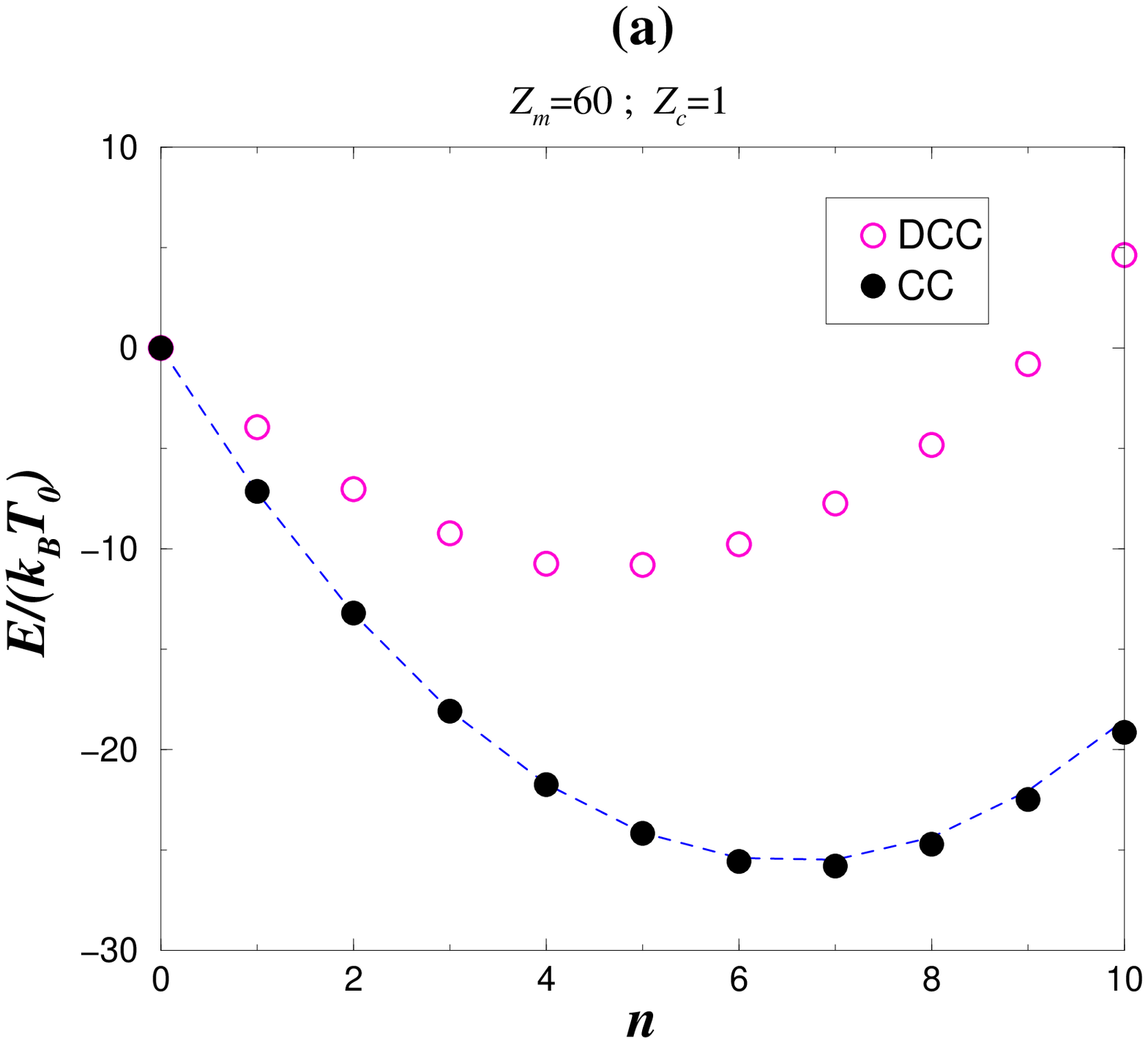}
\includegraphics[width = 6.8 cm]{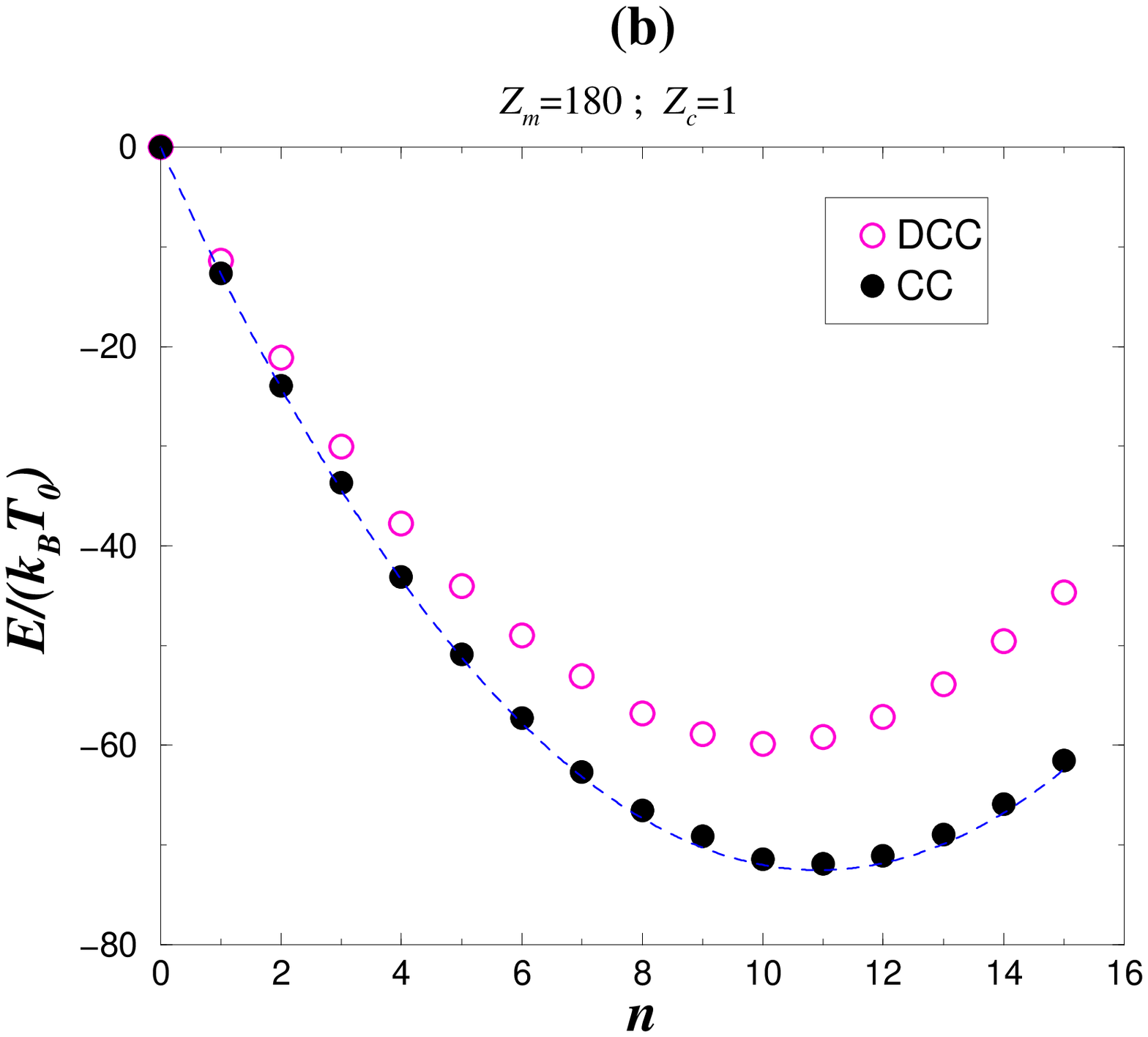}
\hfill
\end{center}
\caption{Total electrostatic energy for \textit{monovalent} counterions ground state
configurations as a function of the number of \textit{overcharging} counterions
$ n $: (a) \textit{$ Z_{m}=60 $} (b) \textit{$ Z_{m}=180 $}.
Overcharge curves were computed for discrete macroion charge distribution (DCC)
and  macroion central charge (CC). The neutral case was chosen as the potential
energy origin. Dashed lines were produced by using Eq. (\ref{Eq.oc}).
For discrete systems (DCC) error bars are smaller than symbols.}
\label{fig.gs-OC-mono}
\end{figure}

Let us first focus on the monovalent symmetric case (-1:+1) where for the neutral
state each DCC site is exactly associated with one counterion as was shown above.
The results in Figs. \ref{fig.gs-OC-mono}(a-b) show that the overcharging
process occurring with a discrete macroion charge distribution is quite different  from
the one obtained with an uniform surface charge distribution. Especially for the smallest
bare charge $ Z_{m}=60 $, the effect of disorder is very important in agreement
with what was already found above for the neutral state in Sec. \ref{sec.GS-neutral-monov}.
The main effects of charge discretization are: (i) the reduction in gain of
energy and (ii) the reduction of maximal (critical) number, $ n^{*} $, of
stabilizing overcharging counterions (corresponding to a minimum in the energy curve).
Both points were thoroughly discussed elsewhere \cite{Messina_EPJE_2001} for
an equivalent symmetric discrete system (-2:+2). It was shown that points (i)
and (ii) can be explained in terms of \textit{ion-dipole} interaction, 
which presently is the main driving force for overcharging. 
 When the overcharging
counterions are present, each of them will essentially interact (attractively)
with its neighboring dipoles (ionic pairs). The attractive ion-dipole interaction
increases with decreasing ion-dipole separation, i. e. increasing macroion charge density
$\sigma_m$.
This explains why the energy gain increases with $Z_m$
{[}compare Fig. \ref{fig.gs-OC-mono}(a) and Fig.
\ref{fig.gs-OC-mono}(b){]}. On the other hand, the repulsion between the counterions
is not fully minimized since they do not adopt the ideal WC structure
that is obtained with a central charge which in turn explains (i) and (ii).
However for high bare charge ($ Z_{m}=180 $) the overcharge curve
obtained with DCC {[}see Fig. \ref{fig.gs-OC-mono}(b){]} approaches the one
from the continuous case as expected for high counterion concentration. This
feature is fully consistent with what was already found in Sec. \ref{sec.GS-neutral-monov},
where it was shown that the order of the counterion structure in the neutral state
(for discrete systems) increases with $ \sigma _{m} $. In other terms, the
WC approach through Eq. (\ref{Eq.oc}) is a good approximation for describing
discrete systems at high $ \sigma _{m} $ since stronger ordering exists.

Common features of overcharging between continuous and discrete systems are
briefly given here. We note that $ n^{*} $ increases with the macroionic
charge $ Z_{m} $. Furthermore, for a given $ n $, the gain in energy 
always increases with $ Z_{m} $. Also, for a given macroionic charge $ Z_{m} $,
the gain in energy between two successive overcharged states is decreasing with
$ n $. Note that at $T=0$ K, the value of $ \epsilon _{r} $ acts only
as a prefactor. All these features are captured by Eq. (\ref{Eq.oc}).

\subsubsection{Multivalent counterions\label{sec.GS-OC-multiv} }

\begin{figure}[t]
\begin{center}
\includegraphics[width = 6.8 cm]{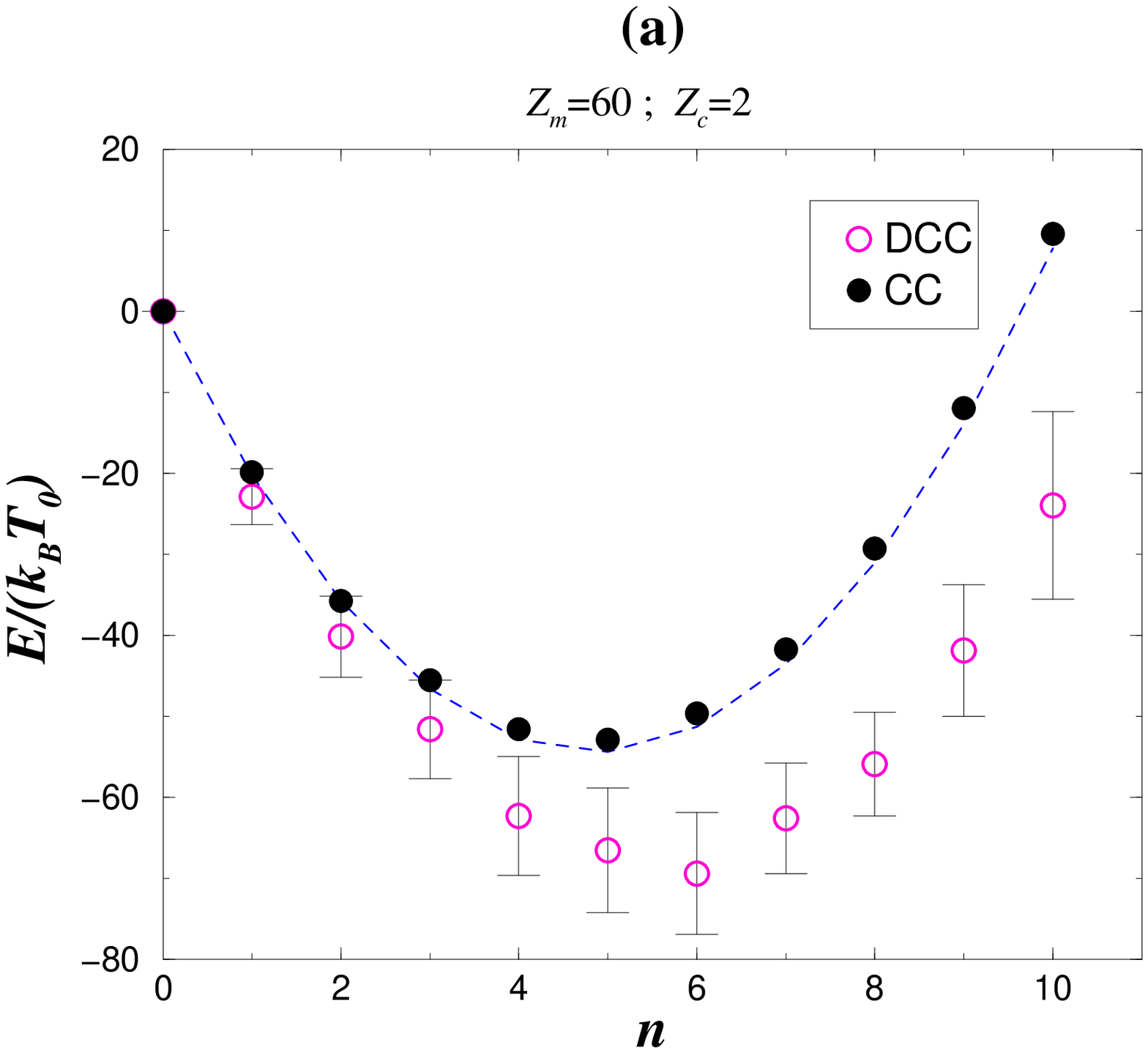}
\includegraphics[width = 6.8 cm]{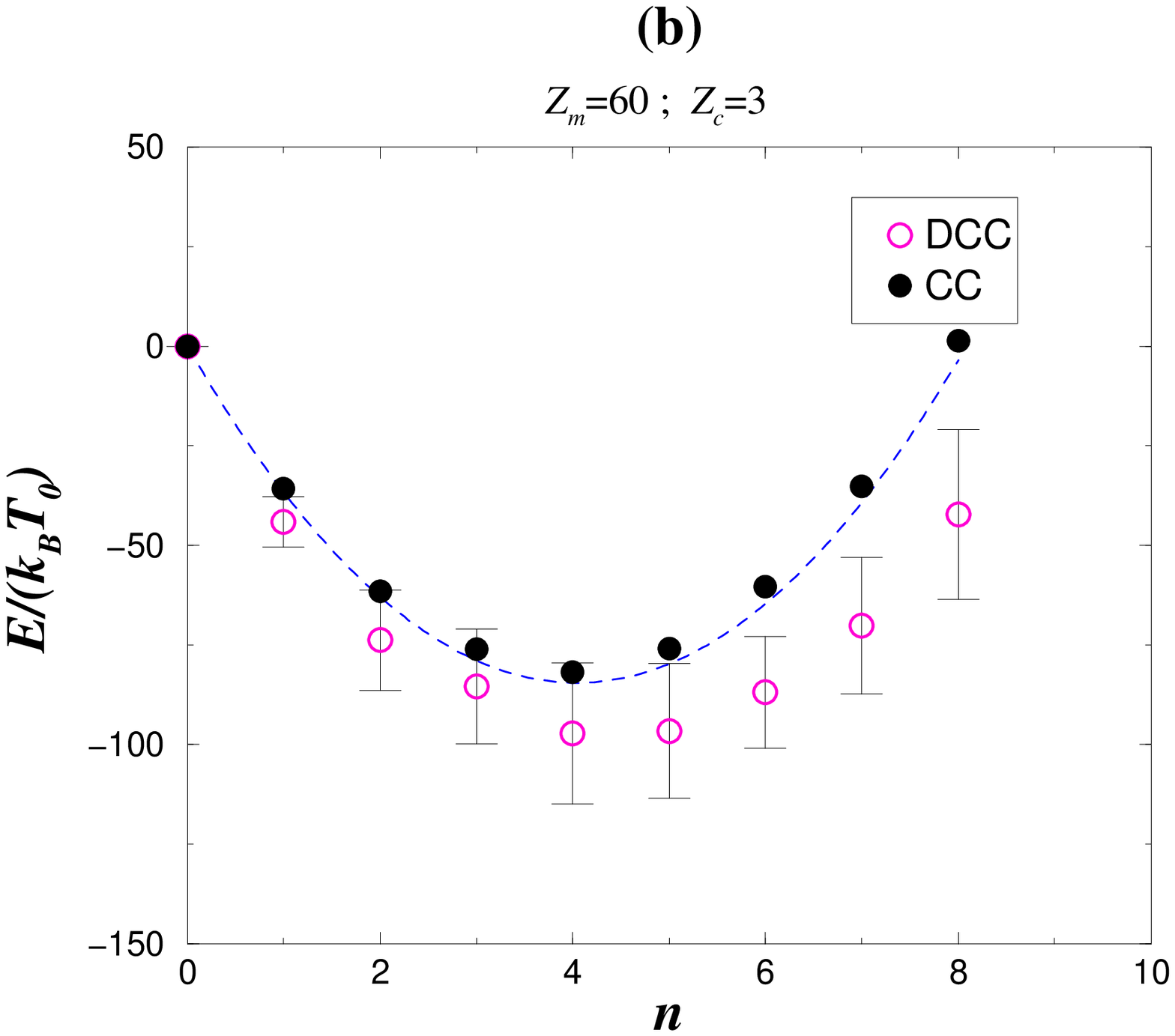}
\includegraphics[width = 6.8 cm]{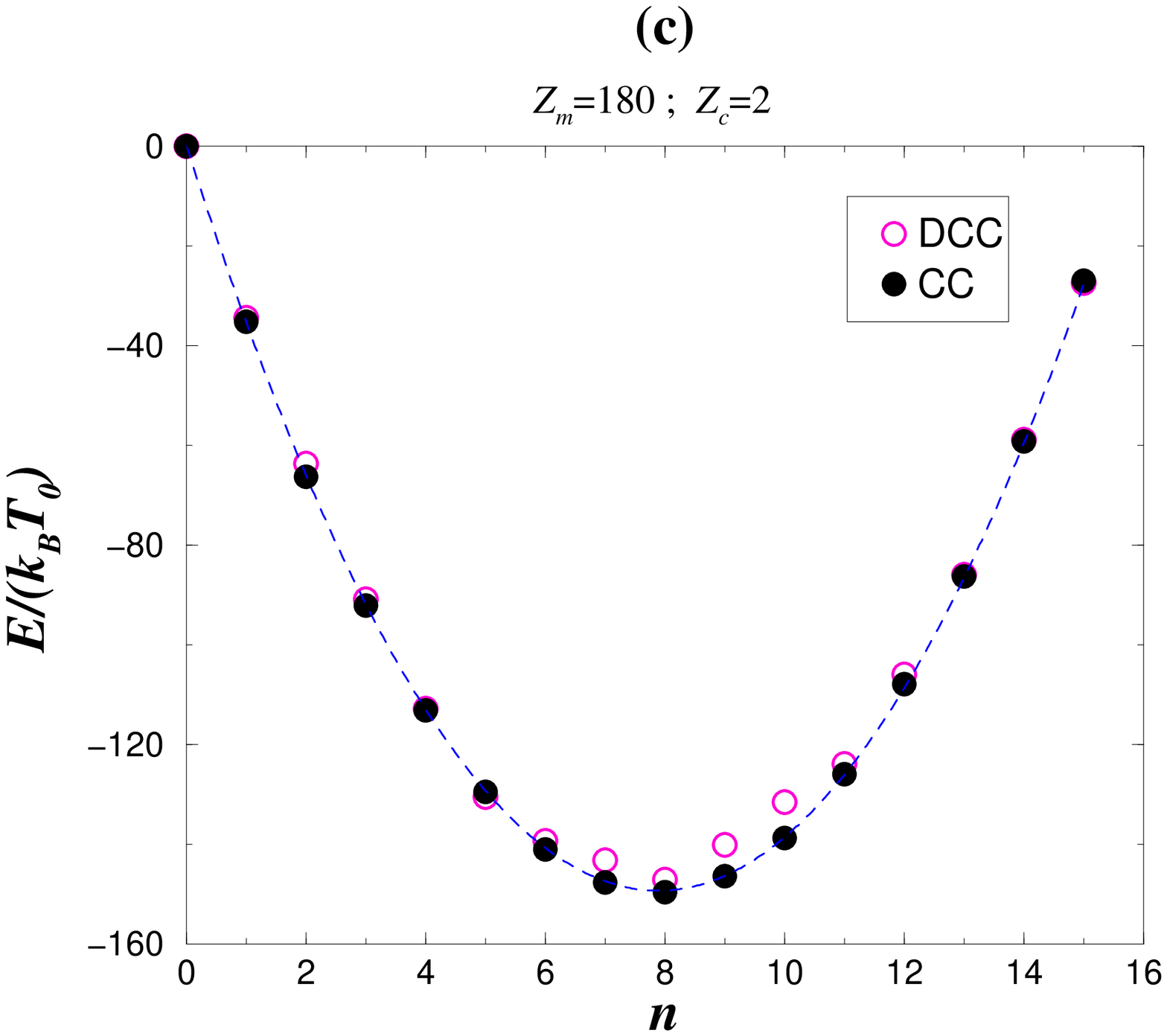}
\includegraphics[width = 6.8 cm]{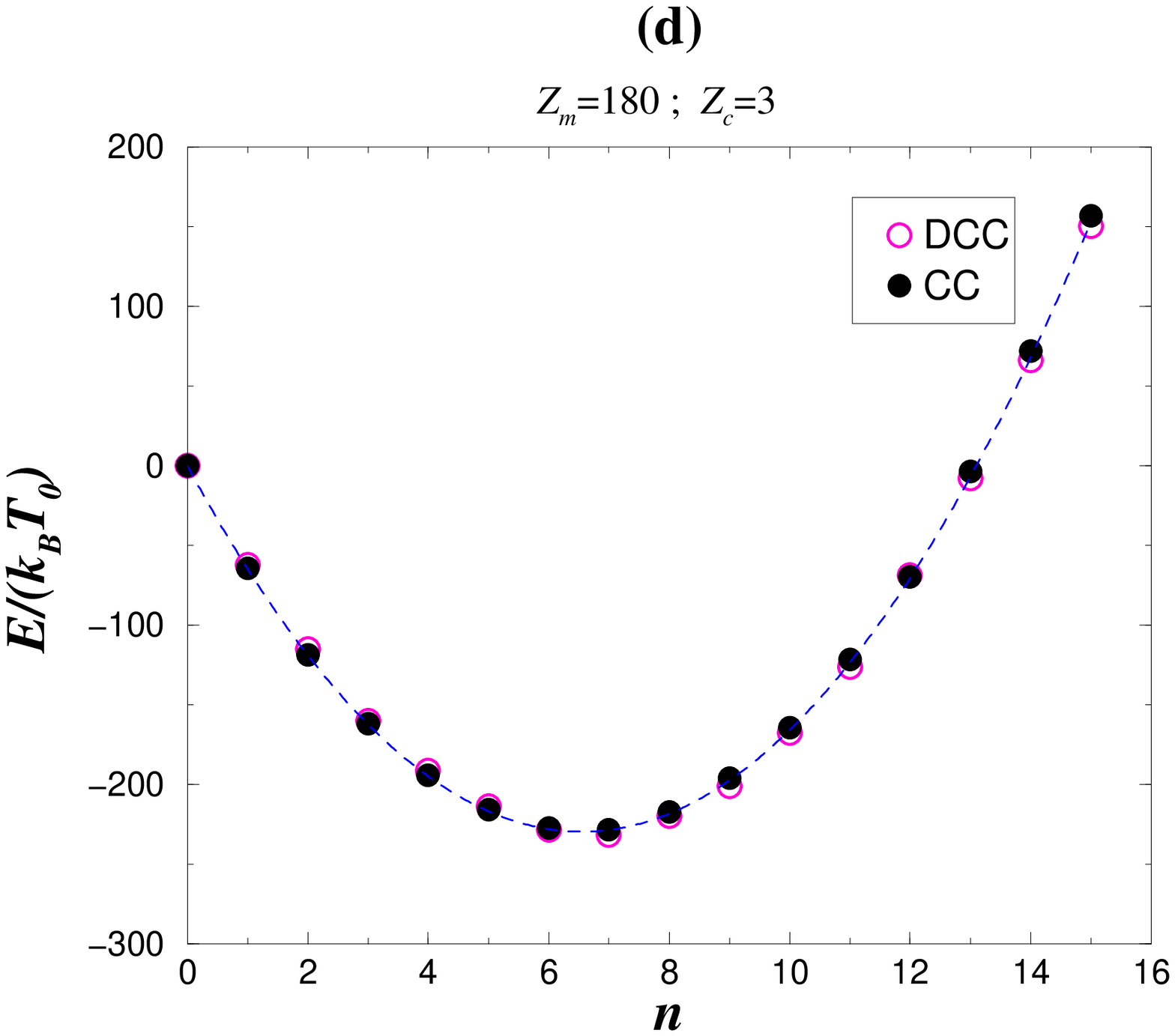}
\hfill
\end{center}
\caption
{Total electrostatic energy for \textit{multivalent} counterions ground state
configurations as a function of the number of \textit{overcharging} counterions
$ n $: (a) \textit{$ Z_{m}=60 $,} $ Z_{c}=2 $
(b) \textit{$ Z_{m}=60 $,} $ Z_{c}=3 $
(c) \textit{$ Z_{m}=180 $,} $ Z_{c}=2 $
(d) \textit{$ Z_{m}=180 $,} $ Z_{c}=3 $.
Overcharge curves were computed for discrete macroion charge distribution (DCC)
and macroion central charge (CC). The neutral case was chosen as the potential
energy origin. Dashed lines were produced by using Eq. (\ref{Eq.oc}).
For discrete systems (DCC) error bars are only indicated when larger than symbols.
}
\label{fig.gs-OC-multi}
\end{figure}

Now we are going to discuss the asymmetric discrete systems $ (-1:+Z_{c}) $
where multivalent counterions are present ($ Z_{c}>1 $). The results of 
figures \ref{fig.gs-OC-multi}(a-d) indicate that the energy gain 
in the overcharging process at fixed
$ Z_{m} $ and $ n $ is higher the higher the counterion valence $ Z_{c} $
for both macroion charge distributions (discrete and continuous). For the continuous case
this can be directly explained in terms of WC concepts {[}i.
e. Eq. (\ref{Eq.oc}){]}. Indeed the main leading term of the correlational
energy $ \Delta E_{n}^{cor} $ in Eq. (\ref{Eq.oc}) scales like

\begin{equation}
\label{eq.oc-Z3-2}
\Delta E_{n}^{cor}\sim -Z^{3/2}_{c}
\end{equation}
for fixed $ n $ and fixed macroion charge $ Z_{m} $, and recalling that
$ N_{c}=Z_{m}/Z_{c} $. Equation (\ref{eq.oc-Z3-2}) quantitatively (qualitatively)
explains why overcharging is stronger with increasing counterion valence $ Z_{c} $
for the continuous (discrete) case.

As far as discrete systems are concerned, the overcharging mechanisms occurring 
with multivalent counterions differ from those occurring with symmetric monovalent 
systems (-1:+1). This is again due to the presence of $ (Z_{m}-Z_{m}/Z_{c}) $ unbound 
DCC sites in the neutral state as discussed in Sec. \ref{sec.GS-neutral-multiv}.
When overcharging comes into play, each overcharging counterion becomes paired
with some\footnote{%
It can be one or more depending on the valence, surface charged density and
the local DCC site arrangement.
} 
of these free DCC sites. 
Figure \ref{fig.gs-OC-multi} shows that  the overcharging with multivalent counterions 
(especially the higher $ Z_{m} $) is significantly less affected by  colloidal charge  
discretization than in the monovalent case (see Fig. \ref{fig.gs-OC-mono}).

For $ Z_{m}=60 $, simulations show that overcharging in the discrete case
can even be stronger than in the continuous case {[}see Figs. \ref{fig.gs-OC-multi}(a) and (b){]}. 
This phenomenon can be qualitatively understood by referring to the very low macroion
surface charge density limit, where the correlation term $ \Delta E_{n}^{cor} $
in Eq. (\ref{Eq.oc}) becomes negligible compared 
to the ionic pairing term $ E_{pin} $ given by Eq. (\ref{eq.Epin}). 
In this limiting situation, the energy gain by overcharging
is approximatively given by $ -nZ_{d}Z_{c}/d_{pin} $ so that
\textit{full} overcharging occurs where each monovalent DCC site is paired with
one multivalent counterion. 

For $ Z_{m}=180 $, the overcharging curves for discrete and continuous distributions 
are almost identical [see  Figs. \ref{fig.gs-OC-multi}(c-d)].
This is consistent with what we already  found
in Sec. \ref{sec.GS-neutral-multiv} for the counterion structure in the neutral state,
where we showed that  $g_c^{ \mathrm {(DCC)}}(r)$
approaches  $g_c^{ \mathrm {(CC)}}(r)$ with increasing $Z_{c}$.
However, the agreement between discrete and continuous cases is even better
for overcharging than for counterion structure {[}see  the corresponding
$g_c(r)$ given in Figs. \ref{fig.gs-CCF-multivalent}(c) and (d){]}.
This is due to the fact that, as previously mentioned, the WC approach [Eq. (\ref{Eq.oc})]
quantifying the energy gain by overcharging is already excellent
for highly short-ranged ordered systems. Generally speaking, all the ordering
mechanisms related in Sec. \ref{sec.GS-neutral} for neutral discrete systems
hold for the overcharging features: \textit{all causes leading to ordering enhance
overcharging}.

\section{Finite temperature\label{sec.Temp}}

In this part, the system is globally \textit{neutral} and is brought to room
temperature $ T_{0} $. We are interested in determining the counterion distribution
as well as the counterion motion within the counterion layer. The cell radius
$ R $ is fixed to $ 40\sigma  $ so that the macroion volume fraction $ f_{m} $
has the \textit{finite} value $ 8\times 10^{-3} $.

\subsection{Strong Coulomb coupling\label{sec.Temp-SCC}}

The Bjerrum length $ l_{B} $ is set to $ 10\sigma  $ as previously in
the ground state study Sec. \ref{sec.GS-neutral}. In this section we consider
two macroion bare charges $ Z_{m} $ (60 and 180) and three counterion valences
$ Z_{c} $ (1, 2 and 3). A typical parameter for describing the Coulomb coupling
strength is the so-called plasma parameter $ \Gamma  $ \cite{Rouzina_JCP_1996}
defined as $\Gamma =l_{B}Z^{2}_{c}/\widetilde{a}_{cc}$.
For our simulation parameters, $ \Gamma  $ ranges from 2.6 (for $ Z_{m}=60 $
and $ Z_{c}=1 $) up to $23.1$ (for $ Z_{m}=180 $ and $ Z_{c}=3 $). Under
these conditions, systems are still highly energy dominated so that at equilibrium
almost all (if not all depending on $ Z_{m} $ and $ Z_{c} $) counterions
lie in the vicinity of the macroion surface (strong
condensation). Therefore for the strong Coulomb coupling regime it is
suitable to focus
on the counterion \textit{surface} properties. In the following sections
we are going to study surface counterion distribution and diffusion.

\subsubsection{Counterion distribution\label{sec.Temp-SCC-coun-distri}}

Like in the ground state analysis, we characterize the counterion distribution
via its surface correlation function. At non zero temperature, correlation functions
are computed by averaging $ \sum _{i\neq j}\delta (r'-r_{i})\delta (r''-r_{j}) $
over 1000 independent equilibrium configurations which are statistically uncorrelated.

\begin{figure}[b]
\begin{center}
\includegraphics[width = 6.8 cm]{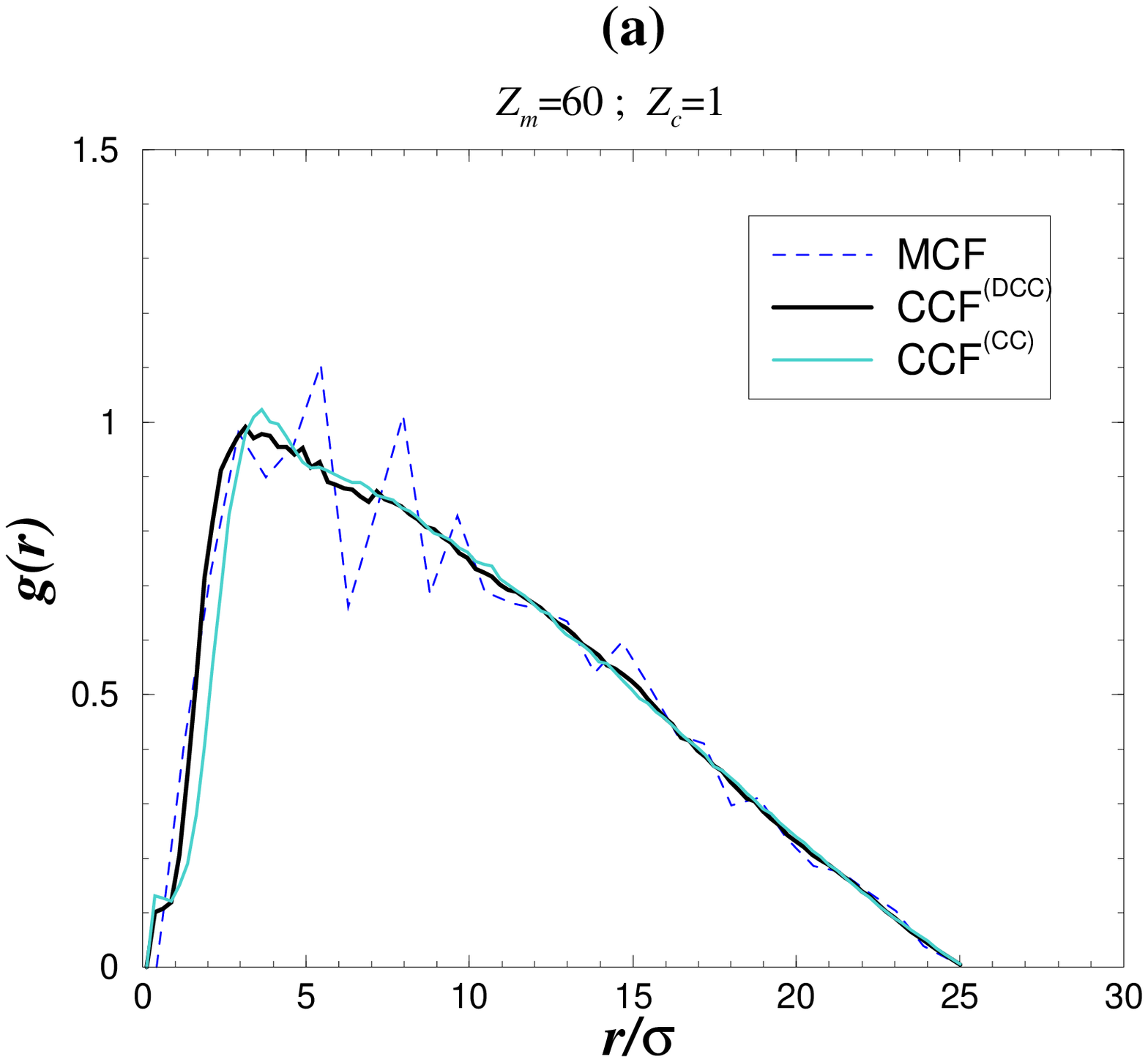}
\includegraphics[width = 6.8 cm]{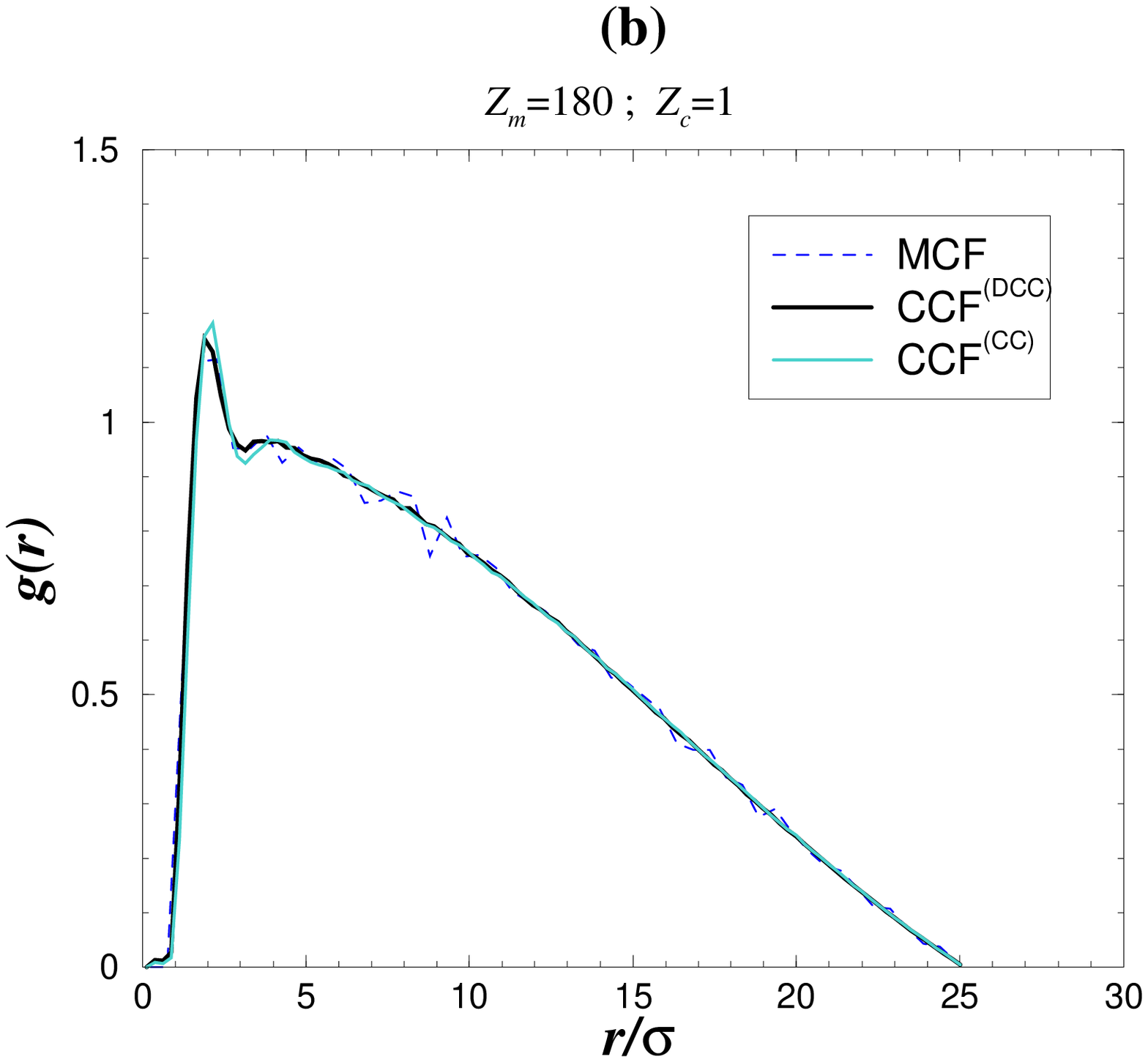}
\end{center}
\caption
{
Surface correlation functions at room temperature $ T_{0} $
for \textit{monovalent} counterions. The two counterion correlation functions
(CCF) $g_c(r)$ are obtained for discrete colloidal charges (DCC) and for the central
charge (CC): (a) \textit{$ Z_{m}=60 $} (b) \textit{$ Z_{m}=180 $}. MCF stands
for $g_m(r)$.
}
\label{fig.Temp-mono-CCF}
\end{figure}

The results for monovalent counterions are depicted in Fig. \ref{fig.Temp-mono-CCF}(a)
and Fig. \ref{fig.Temp-mono-CCF}(b) for $ Z_{m}=60 $ and $ Z_{m}=180 $ respectively.
For both charges $ Z_{m} $ the counterion distributions are weakly affected
by charge discretization and $g_c^{ \mathrm {(DCC)}}(r)$ and $g_c^{ \mathrm {(CC)}}(r)$
are almost identical. 
A closer look on Fig. \ref{fig.Temp-mono-CCF} reveals that the agreement between discrete
and continuous distributions is even better for high macroion charge density ($Z_{m}=180$)
as expected. In fact for monovalent systems the pinning term $ E_{pin} $
has its lowest magnitude so that, for sufficiently high $\sigma_m$, the \textit{fluctuating}
intra-dipole separation becomes comparable to the inter-dipole separation and
discretization effects (i. e. pinning) are canceled. These pinning and unpinning
aspects will be addressed in more details in the next section \ref{sec.Temp-SCC-diff}.
As expected, the counterion positional order for discrete and continuous cases
is much weaker at room temperature than in the ground state case (compare Fig.
\ref{fig.Temp-mono-CCF} and Fig. \ref{fig.gs-CCF-monovalent}).

\begin{figure}[t]
\begin{center}
\includegraphics[width = 6.8 cm]{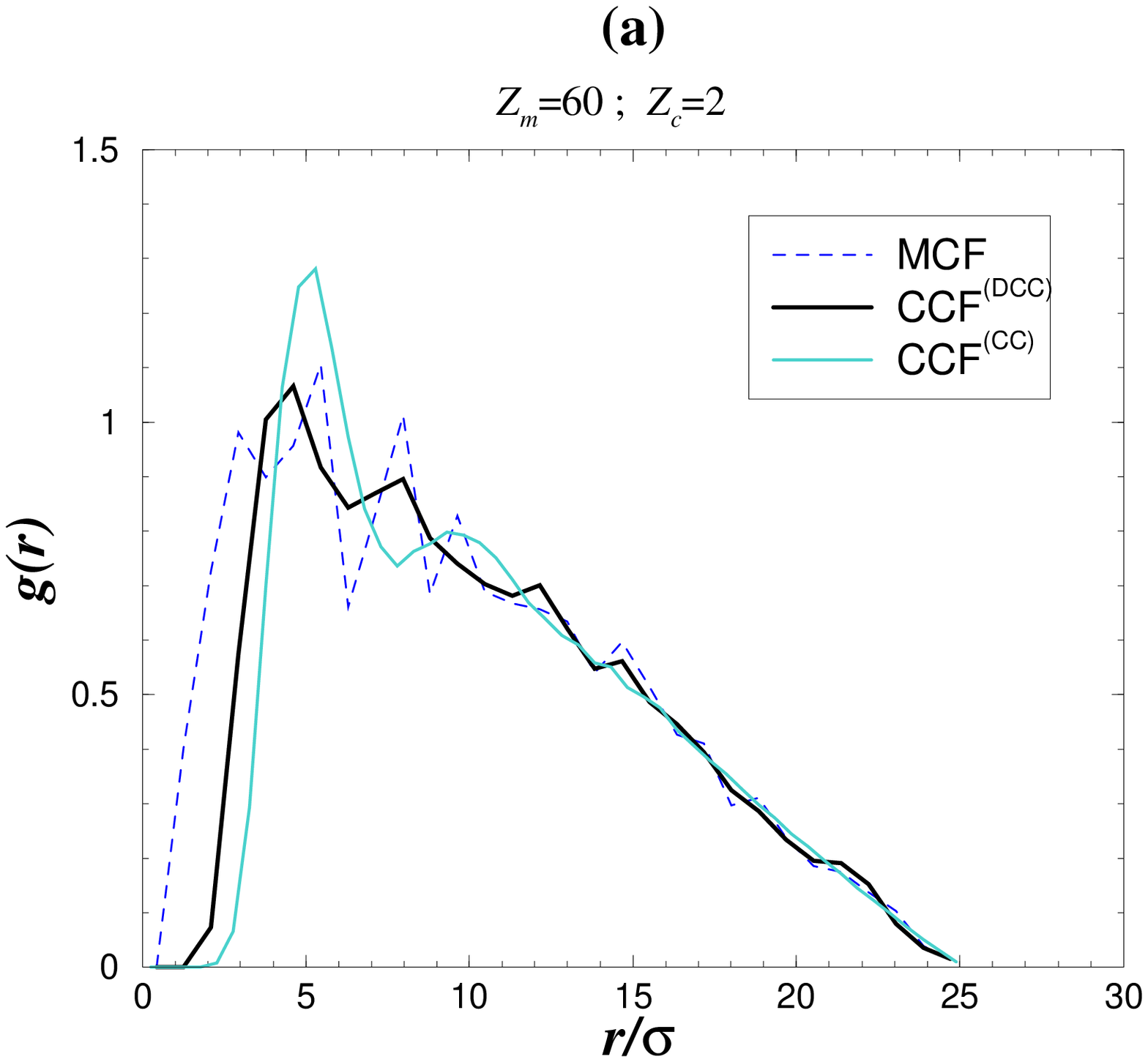}
\includegraphics[width = 6.8 cm]{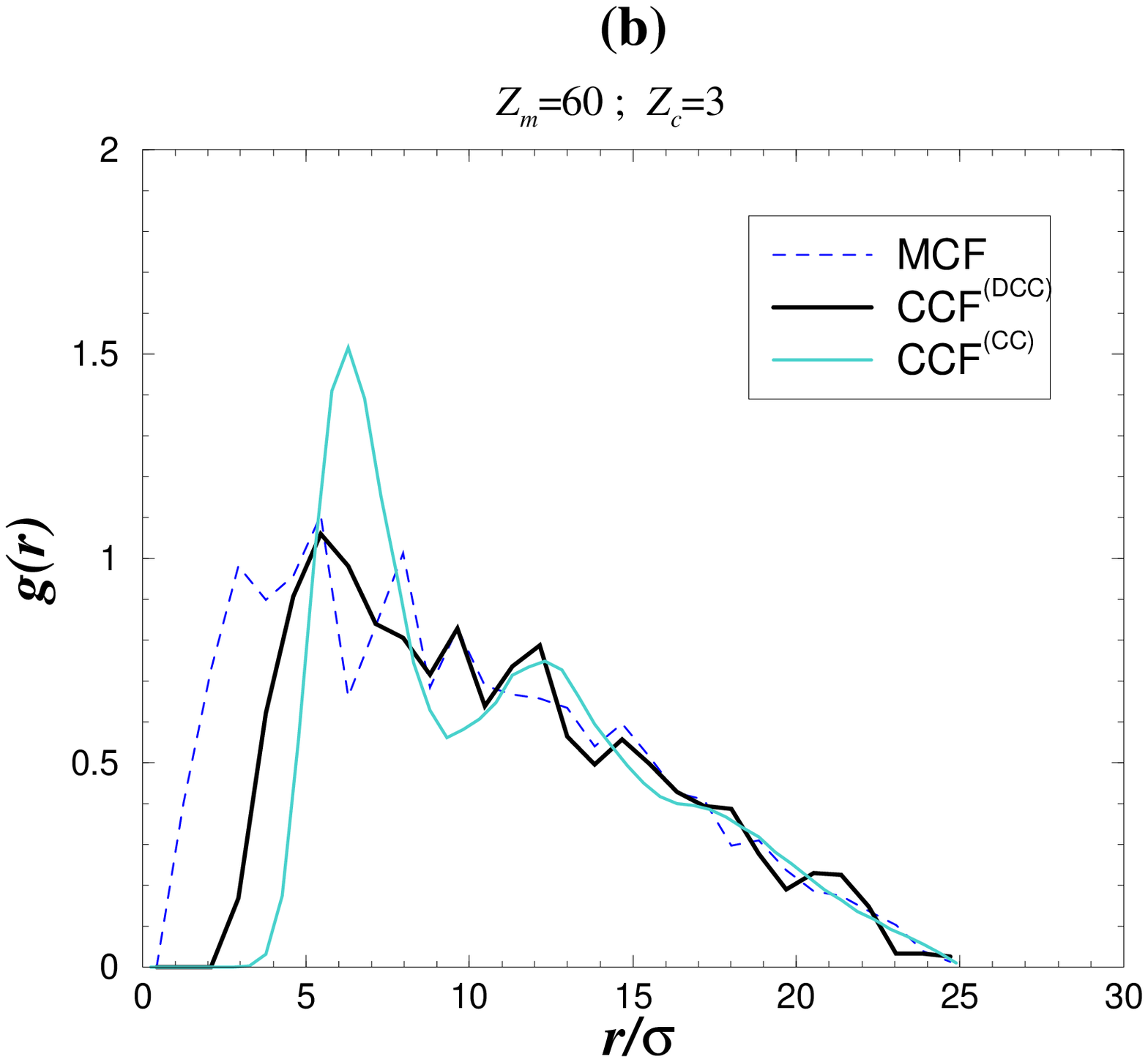}
\includegraphics[width = 6.8 cm]{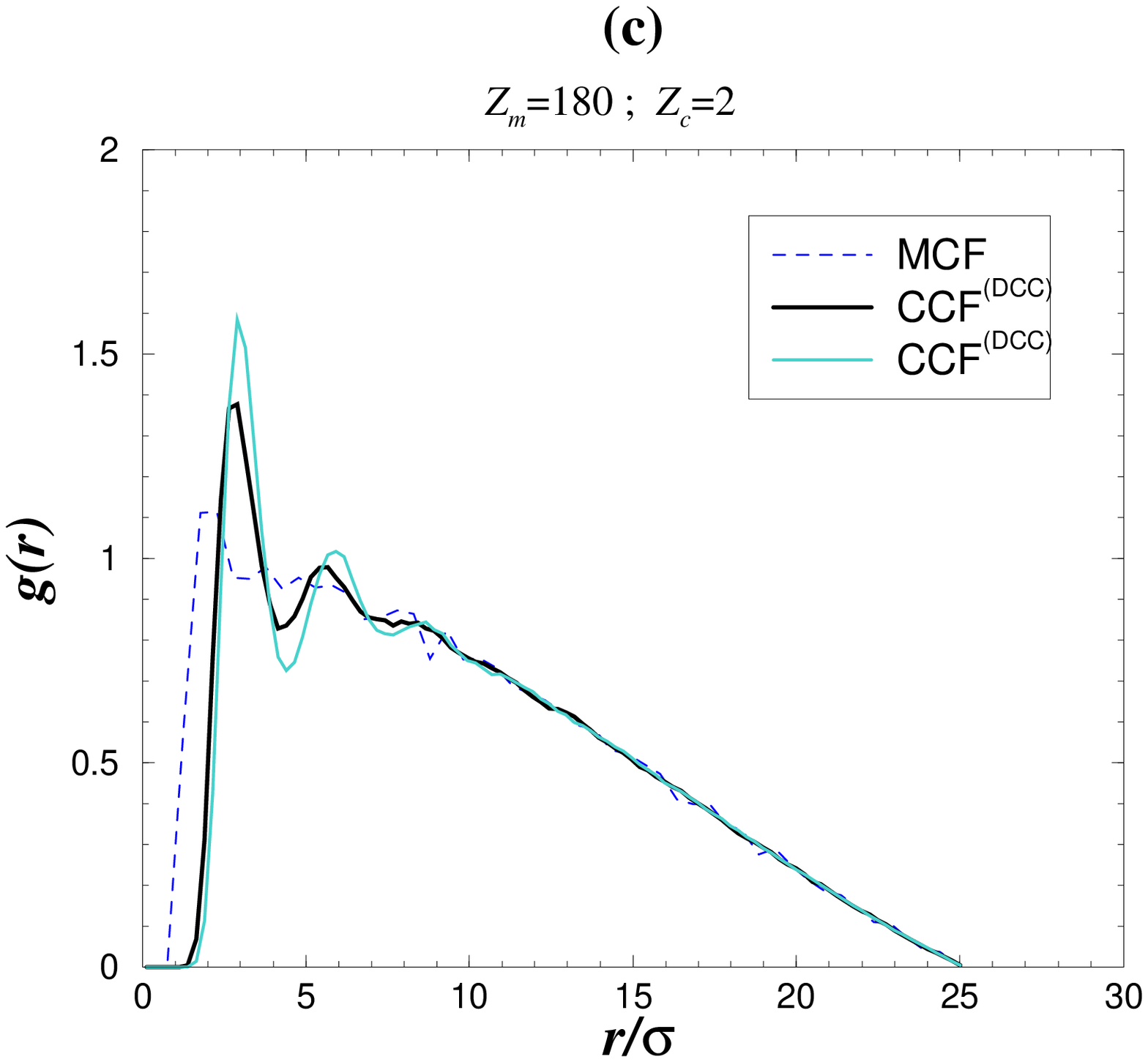}
\includegraphics[width = 6.8 cm]{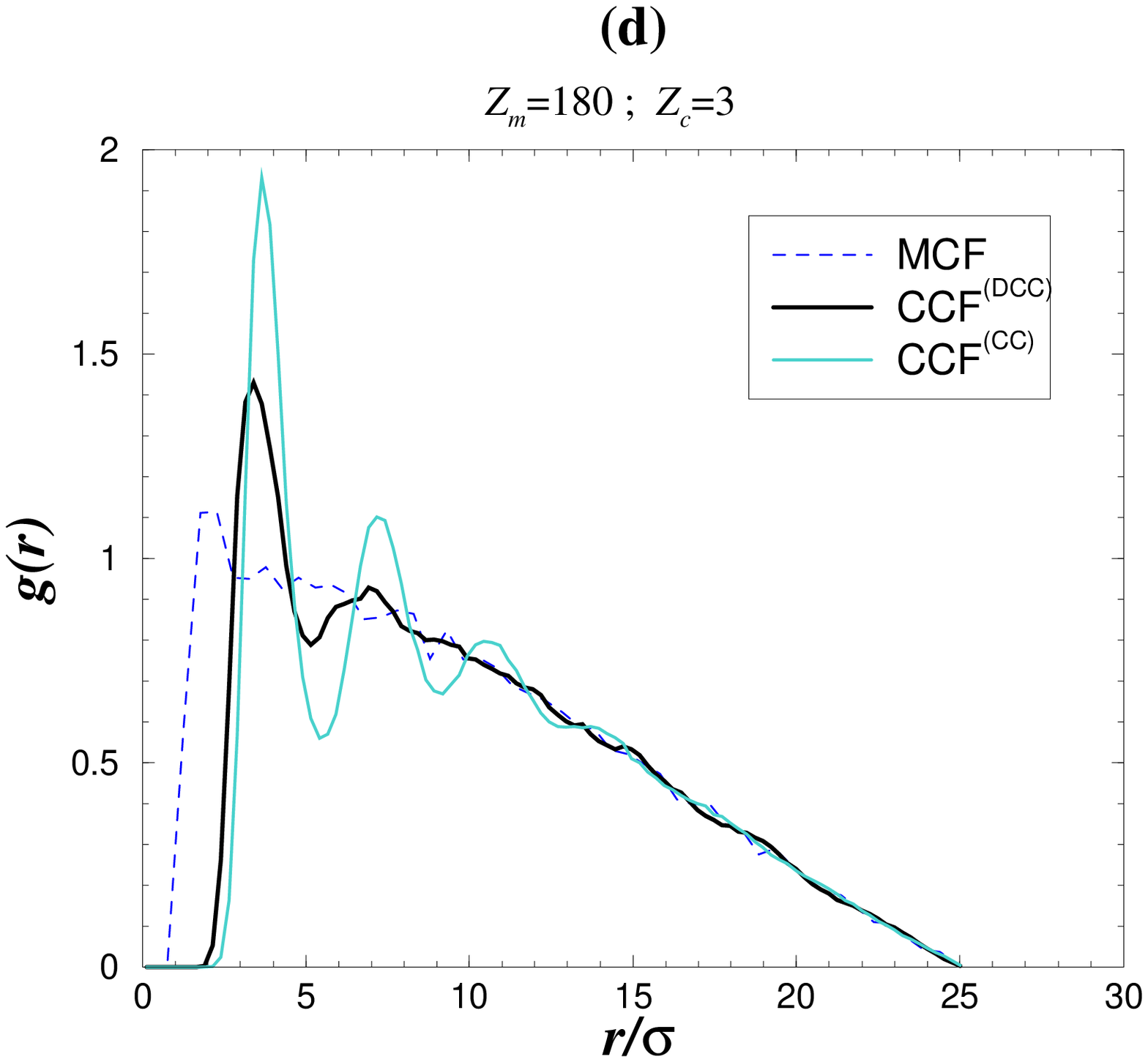}
\end{center}
\caption
{
Surface correlation functions at room temperature $ T_{0} $
for \textit{multivalent} counterions. The two counterion correlation functions
(CCF) $g_c(r)$ are obtained for discrete colloidal charges (DCC) and for the central
charge (CC): (a) \textit{$ Z_{m}=60 $,} $ Z_{c}=2 $
(b) \textit{$ Z_{m}=60 $}, $ Z_{c}=3 $
(c) \textit{$ Z_{m}=180 $,} $ Z_{c}=2 $
(d) \textit{$ Z_{m}=180 $,} $ Z_{c}=3 $.
MCF stands for $g_m(r)$.
}
\label{fig.Temp-multi-CCF}
\end{figure}

The results for multivalent counterions are depicted in Fig. \ref{fig.Temp-multi-CCF}.
We now find that the counterion distributions are strongly affected
by charge discretization, and especially the higher $ Z_{c} $. This is in contrast
with what was found in the ground state analysis Sec. \ref{sec.GS-neutral-multiv}
where no counterion motion occurs. This effect is of course due to the pinning
(inhibition of large counterion motion) which is proportional to $ Z_{c} $.

Note that all the statements above hold for the particular finite temperature
$ T_{0} $. However the effect of finite temperature discussed here should
hold, at least qualitatively, for a large temperature range. For very low temperature
one should recover all ground state properties mentioned in Sec. \ref{sec.GS-neutral}.

\subsubsection{Surface diffusion\label{sec.Temp-SCC-diff}}

This section is devoted to answer the following question: do the counterions
only oscillate around the DCC sites or do they have also a large translational
motion over the sphere?

To study this problem we introduce the following observable:

\begin{equation}
\label{eq.MSD}
\Delta x^{2}(t,t_{0})=\frac{1}{t-t_{0}}\int ^{t}_{t_{0}}dt^{'}[x(t^{'})-x(t_{0})]^{2} ,
\end{equation}
which is referred as the \textit{mean square displacement} (MSD), where $ x(t_{0}) $
represents the position of a given counterion at time $ t=t_{0} $ (at equilibrium)
and $ x(t,t_{0}) $ is its position at later time \textit{t}. 
All particles lying within a distance $ 9.2\sigma  $ from the macroion center
are radially projected on the macroion surface of radius $ a=8\sigma  $ to
give $ x(t,t_{0}) $. The root mean square displacement (RMSD) $ \Delta x(t,t_{0}) $
is defined as

\begin{equation}
\label{eq.RMSD}
\Delta x(t,t_{0})=\sqrt{\Delta x^{2}(t,t_{0})}.
\end{equation}
Like for the surface correlation function, the RMSD is measured on the spherical
surface (arc length) and it has a natural upper limit $ \pi a $. For the
case of free counterions (i. e. macroion central charge without pinning) the RMSD
$ \Delta x_{free} $ reads

\begin{equation}
\label{eq.RMSD-free}
\Delta x_{free}=a\sqrt{\frac{\pi ^{2}-4}{2}}\approx 13.7\sigma .
\end{equation}
This quantity $ \Delta x_{free} $ will be useful to refer to the {}``unpinned{}''
state.

\begin{figure}[b]
\begin{center}
\includegraphics[width = 6.8 cm]{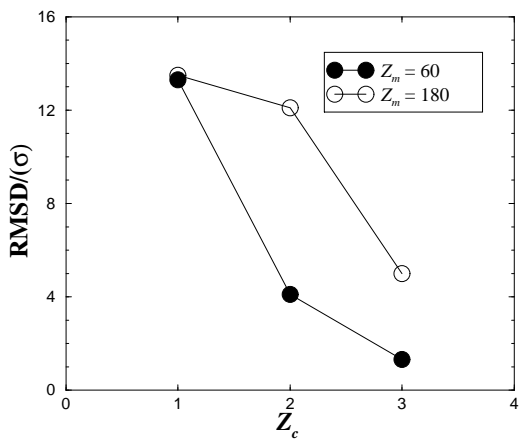}
\hfill
\end{center}
\caption{
Root mean square displacement (RMSD) as a function of counterion
valence $ Z_{c} $ for $ Z_{m}=60 $ and $ Z_{m}=180 $.
Errors are smaller than symbols.
}
\label{fig.surface diffusion}
\end{figure}

The results for discrete systems are sketched in Fig. \ref{fig.surface diffusion}
for $ Z_{m}=60 $ and $ Z_{m}=180 $. 
Monovalent counterions
are free to move over the macroion surface for both bare charges $ Z_{m} $
considered here. Moreover, our simulation data show that 
the counterions gradually become pinned with increasing $ Z_{c} $. All these
features are captured by the $ Z_{c} $ dependency of the pinning term $ E_{pin} $.
For multivalent counterions, the degree of pinning increases with decreasing
$ Z_{m} $. This is due to the fact that the discrete charges
get closer from each other by increasing $ Z_{m} $
so that a counterion jump from site to site is energetically less demanding.
For the continuous case, we have checked that for the same parameters counterions
always have a large lateral motion and move all over the sphere.

\subsection{Moderate Coulomb coupling\label{sec.Temp-WCC}}

In this last part, the Bjerrum length corresponds to that of $water$  at
room temperature ($ l_{B}=2\sigma =7.14 $ \AA).\ For this moderate Coulomb
coupling counterions occupy $all$ the cell volume. Clearly, the probability of
finding counterions plainly outside the macroion surface is \textit{no more}
negligible (in contrast with the strong Coulomb coupling). The target quantity is
the fraction $ P(r) $ of counterions lying within a  distance $r$ from the
macroion center and is defined as

\begin{equation}
\label{eq.P_r}
P(r)=N(r)/N_{c}
\end{equation}
with

\begin{equation}
\label{eq.N_r}
N(r)=\int ^{r}_{r_{0}}4\pi r^{2}_ic_{v}(r_i)dr_i,
\end{equation}
where $ c_{v}(r) $ is the profile of the  $volume$ counterion concentration  and 
$ N(r) $ is the so-called integrated charge.
\begin{figure}[b]
\begin{center}
\includegraphics[width = 6.8 cm]{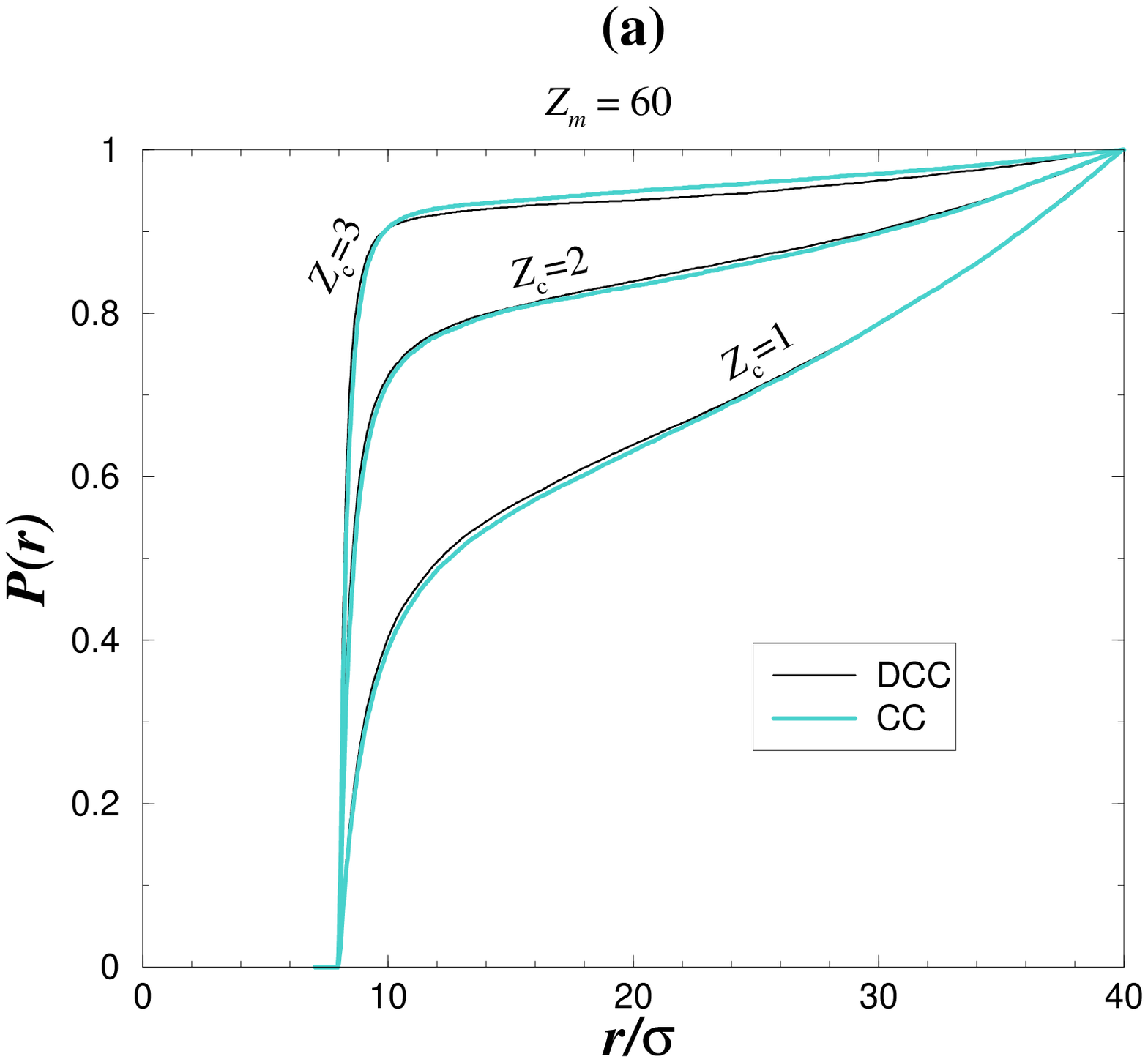}
\includegraphics[width = 6.8 cm]{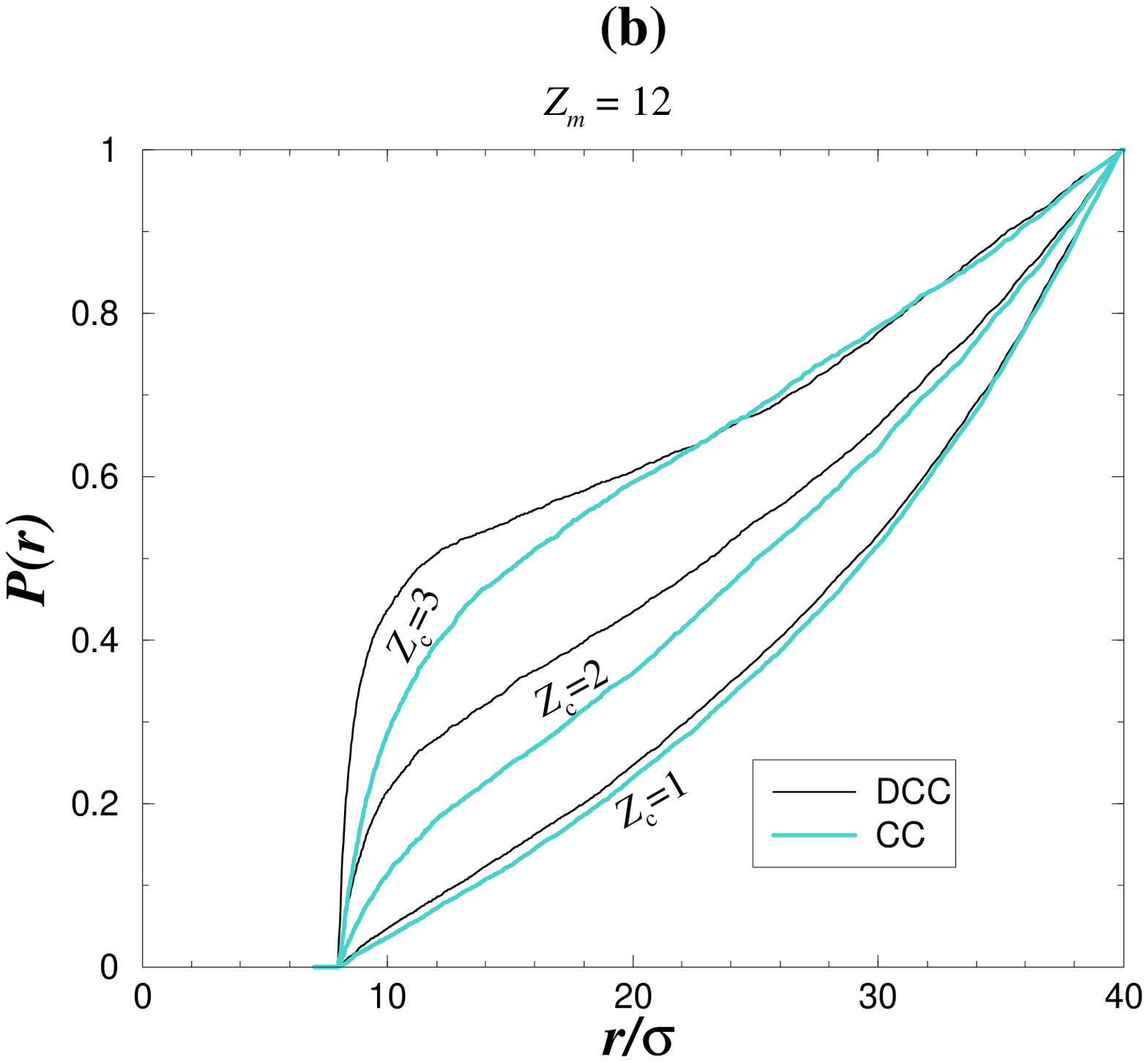}
\hfill
\end{center}
\caption
{
Counterion fraction within a distance $ r $ from the macroion
center for different counterion valence $ Z_{c} $. (a) $ Z_{m}=60$
(b) $Z_{m}=12$. Data were obtained for discrete
macroion charge distribution (DCC) and macroion central charge (CC).
}
\label{fig.volume_distribution}
\end{figure}

The results for $ Z_{m}=60 $ and $ Z_{m}=12 $ are sketched in Fig. \ref{fig.volume_distribution}(a)
and Fig. \ref{fig.volume_distribution}(b) respectively. 
For the highest charge, Fig. \ref{fig.volume_distribution}(a)
shows that discretization effects are canceled for any counterion valence. On the other hand,
for the small charge density case, Fig. \ref{fig.volume_distribution}(b) shows
that discretization effects become important for multivalent counterions. In
the present situation, the Coulomb coupling is five times less important than
in the strong coupling case studied in Sec. \ref{sec.Temp-SCC}. Therefore pinning
effects  can only be noticeable for sufficiently low $ \sigma _{m} $
(here $ Z_{m}=12 $) and multivalent counterions.

\section{Conclusion \label{sec.Conclusion}}

We have performed MD simulations within the framework of the primitive model
to study the coupled effects of macroion charge discretization and counterion
valence. The macroion bare charge is carried by monovalent microions randomly
distributed over the colloidal surface. Different correlational regimes were
considered: (i) ground state and (ii) finite temperature.

Concerning the ground state analysis, we were interested in the counterion
structure in the neutral state and the overcharging phenomenon. We demonstrated
that the order in the surface counterion structure  (disorder in counterion structure 
induced by the discrete random macroion charge distribution) is increased 
(decreased) by increasing macroion surface charge density $ \sigma _{m} $ and/or 
counterion valence $ Z_{c} $. For monovalent counterions, we showed that 
the ratio between the intra-ion pair (made up of a discrete colloidal surface
ion and a counterion) distance and the mean distance between ion pairs
is a fundamental quantity to describe counterion ordering. When overcharge comes
into play similar effects occur. More precisely, for sufficiently high charge density
$ \sigma_{m} $ the overcharging with monovalent
as well as multivalent counterions is \textit{quantitatively} the same as the
one obtained in the continuous case. For low $ \sigma_{m} $,
the overcharging with multivalent counterions can even be stronger in the discrete case
than in the continuous case counterions. In contrast, for monovalent counterions
overcharging is always weaker
than in the continuous case but approaches the latter with increasing $ \sigma _{m} $.

In the finite temperature case, strong and moderate Coulomb couplings
were addressed. In the strong Coulomb coupling, we showed that counterion pinning
is very weak for monovalent counterions but it increases with increasing $Z_{c}$ 
and decreasing $ \sigma _{m} $. This
involves an increasing disorder in the surface counterion structure with
increasing $ Z_{c} $ and decreasing $ \sigma _{m} $. 
In the moderate Coulomb coupling corresponding to
an aqueous situation, the $volume$ counterion distribution is only affected for low
$ \sigma _{m} $ and multivalent counterions.

A future work will address the presence of added salts as well as the case of
two interacting macroions.

\ack

I thank C. Holm, A. Johner, K. Kremer and B. Shklovskii for helpful discussions.
This work is supported by
\textit{Laboratoires Europ\'{e}ens Associ\'{e}s} (LEA).


\begin{thebibliography}{10}
\expandafter\ifx\csname url\endcsname\relax
  \def\url#1{\texttt{#1}}\fi
\expandafter\ifx\csname urlprefix\endcsname\relax\def\urlprefix{URL }\fi

\bibitem{Isralachvili_1992}
J.~Israelachvili, Intermolecular and Surface Forces, Academic, London, 1992.

\bibitem{Evans_book_1999}
D.~F. Evans and  H.~Wennerstr\"om, The Colloidal Domain, Wiley-VCH, New York, 1999.

\bibitem{Hill_book_1960}
T.~L. Hill, Statistical mechanics, Addison-Wesley, Reading, Mass., 1960.

\bibitem{Wennerstroem_JCP_1982}
H.~Wennerstr\"om, B.~J\"onsson and P.~Linse, J. Chem. Phys. 76 (1982) 4665.

\bibitem{Perel_Physica_1999}
V.~Perel and B.~Shklovskii, Physica 274A (1999) 446.

\bibitem{Shklowskii_PRE_1999b}
B.~Shklovskii, Phys. Rev. E 60 (1999) 5802.

\bibitem{Mateescu_EPL_1999}
E.~M. Mateescu, C.~Jeppesen and P.~Pincus, Europhys. Lett. 46 (1999) 493.

\bibitem{Joanny_EPJB_1999}
J.~F. Joanny, Europ. J. Phys. B 9 (1999) 117.

\bibitem{Sens_PRL_1999}
E.~Gurovitch and P.~Sens, Phys. Rev. Lett. 82 (1999) 339.

\bibitem{Marcelo_PRE_RapCom1999}
M.~Lozada-Cassou, E.~Gonz\'alez-Tovar and W.~Olivares, Phys. Rev. E 60 (1999)
  R17.

\bibitem{Deserno_Macromol_2000}
M.~Deserno, C.~Holm and S.~May, Macromolecules 33 (2000) 199.

\bibitem{Messina_PRL_2000}
R.~Messina, C.~Holm and K.~Kremer, Phys. Rev. Lett. 85 (2000) 872.

\bibitem{Messina_EPL_2000}
R.~Messina, C.~Holm and K.~Kremer, Europhys. Lett. 51 (2000) 461.

\bibitem{Nguyen_PRL_2000}
T.~T. Nguyen, A.~Y. Grosberg and B.~I. Shklovskii, Phys. Rev. Lett. 85 (2000)
  1568.

\bibitem{Nguyen_JCP_2000}
T.~T. Nguyen, A.~Y. Grosberg and B.~I. Shklovskii, J. Chem. Phys. 113 (2000)
  1110.

\bibitem{Messina_EPJE_2001}
R.~Messina, C.~Holm and K.~Kremer, Eur. Phys. J. E 4 (2001) 363.

\bibitem{Messina_PRE_2001}
R.~Messina, C.~Holm and K.~Kremer, Phys. Rev. E 64 (2001) 021405.

\bibitem{Spalla_JCP_1991}
O.~Spalla and L.~Belloni, J. Chem. Phys. 95 (1991) 7689.

\bibitem{Bhattacharjee_Lang_1998}
S.~Bhattacharjee, C.~H. Ko and M.~Elimelech, Langmuir 14 (1998) 3365.

\bibitem{Schmitz_1999}
K.~S. Schmitz, Langmuir 15 (1999) 2854.

\bibitem{Bonsall_PRB_1977}
L.~Bonsall and A.~A. Maradudin, Phys. Rev. B 15 (1977) 1959.

\bibitem{Rouzina_JCP_1996}
I.~Rouzina and V.~A. Bloomfield, J. Chem. Phys. 100 (1996) 9977.

\end{thebibliography}

\newpage














%
\end{document}